\documentclass[useAMS,usenatbib]{mnras}
\usepackage{graphicx,natbib,amssymb,amsmath,stmaryrd,makecell,ulem}
\usepackage{txfonts}
\usepackage{xcolor}
\usepackage{hyperref}

\def \ksz{{kinetic SZ} }
\def \tsz{{thermal SZ} }

\newcommand{\beq}{\begin{equation}}   %

\newcommand{\eeq}{\end{equation}}   %

\newcommand{\beqa}{\begin{eqnarray}}   %

\newcommand{\eeqa}{\end{eqnarray}}   %

\newcommand{\beal}{\begin{align}}

\newcommand{\enal}{\end{align}}

\newcommand{\bspl}{\begin{split}}

\newcommand{\espl}{\end{split}}

\newcommand{\bsub}{\begin{subequations}}

\newcommand{\esub}{\end{subequations}}

\newcommand{\bmulti}{\begin{multline}}   %

\newcommand{\beqm}{\begin{mathletters}}   %

\newcommand{\eeqm}{\end{mathletters}}   %

\title[Quasar]{
Revisiting secondary CMB distortions due to kinetic Sunyaev-Zel'dovich effect from quasar bubbles before reionization}

\begin{document}
\author[Acharya et al.]
{Sandeep Kumar Acharya$^1$\thanks{E-mail:sandeep.acharya@manchester.ac.uk},
and Subhabrata Majumdar$^2$\thanks{E-mail:subha@tifr.res.in}
\\
$^1$Jodrell Bank Centre for Astrophysics, School of Physics and Astronomy, The University of Manchester, Manchester M13 9PL, U.K.
\\
$^{2}$Department of Theoretical Physics, Tata Institute of 
Fundamental Research, Mumbai 400005, India\\
}

\maketitle

\begin{abstract}
We discuss the secondary cosmic microwave background (CMB) anisotropy due to kinetic Sunyaev-Zel'dovich (kSZ) effect from ionized bubbles around individual quasars prior to the reionization of the Universe. The bubbles create local ionization modulations which move with the large scale structure linear bulk flow and act as sources for kSZ. We improve upon previous calculations of this effect, using a halo model based description of quasar abundance, and find that the kSZ distortion power spectrum, $C_\ell$, from the bubbles to be sub-dominant to kSZ from patchy reionization driven by galaxies. However, the shape of the two $C_\ell$'s are very different with the quasar bubble $C_\ell$ having a peak at $\ell \approx 500-700$ whereas the $C_\ell$ due to patchy reionization flattening out at $\ell > 1000$ thus making it plausible to separate the two using $C_\ell$ template-fitting in a future survey like CMB-HD. Next, we look at the imprint of a single quasar bubble on the CMB and show that it can be detected in a high-resolution, ambitious effort like CMB-HD. A detection of a high redshift quasar bubble in the CMB would carry complimentary information to its detection in HI or Lyman-$\alpha$ and a joint analysis can be used to break parameter degeneracies.

\end{abstract}

\section{Introduction}

Cosmological parameters have now been measured to exquisite precision from primary Cosmic Microwave Background (CMB)  anisotropies by the satellite missions WMAP \& Planck   \citep{WMAP2011,Pl2018}. Ground based arcminute resolution experiments such as the Atacama Cosmology Telescope (ACT) \citep{ACT2020} and South Pole Telescope (SPT) \citep{SPTksz2021} have further enriched our understanding of the universe with precise measurement of both primary and secondary CMB anisotropies. While the primary anisotropies are generated at the surface of last scattering ($z\approx 10^3$), these secondary anisotropies are produced by the interaction of the  CMB photons, travelling from the surface of last scattering ($z\approx 10^3$), with the intervening matter at lower redshifts  (see \citep{AMS2008} for a review).

The secondary temperature anisotropies of the CMB can arise from a multitude of effects. They can be of gravitational origin, for example: (i) The Integrated Sachs-Wolfe (late-ISW) effect \citep{SW1967} due to the gravitational redshift/blueshift of CMB photons travelling through a changing, time-dependent gravitational potential, (ii) the Rees-Sciama effect \citep{RS1968} due to the CMB photons passing through evolving, non-linear gravitational potential of largest structures in the universe, and (iii) the gravitational lensing \citep{LC2006} due to the deflection of the CMB photons by the largest structure of the universe. They can also arise from photons with hot baryons, (iv) the thermal Sunyaev-Zeldovich (tSZ) effect \citep{ZS1969} due to interaction of the CMB photons by hot, thermal or non-thermal electrons, and (v) the kinetic SZ effect (kSZ) \citep{SZ1980,MF2002} due to doppler boosting by a bulk flow of electrons. Secondary CMB polarization can also be generated by the scattering of CMB quadrupole with free electrons at the epoch of reionization, as well as from leakage of primary E-mode power to B-mode power due to lensing \citep{ZS1997,H2000}.  
  
Within the different types of secondary CMB distortions, the \ksz is particularly useful as a probe of the high redshift universe. 
The kSZ, as mentioned earlier, is due to scattering from moving electrons  which gives rise to a doppler shift due to their velocities.
Since velocity is a vector, for a distribution of electrons with random velocities, the signal averages out to zero. However, overdensities (like galaxy clusters) can have a bulk velocity as part of the cosmic flow. This gives rise to an overall kSZ effect from these structures. Unfortunately, the spectral signature of the kSZ is that of a black body which makes it difficult to distinguish from the primary CMB based solely on their spectral information. At low redshifts, ($z\lesssim 2$), the \tsz effect is typically dominant over the \ksz by an order of magnitude (or more) for  massive objects \citep{SBDHHLOT2010}, like galaxy clusters, which are abundant at relatively low redshifts. In contrast, the \ksz can contain a treasure trove of cosmological information at $z\gtrsim 2$. 

The \ksz is a complementary probe of cosmology, compared to growth of density probes, since it depends on velocities that are sourced by the growth of overdensities. The magnitude of kSZ signal is given by $y_{kSZ}=\frac{v}{c}\tau$ where $v$ is the cosmological bulk velocity of the ionized source responsible for scattering the CMB photons and $\tau$ is the optical depth for the CMB photons passing through the ionized medium. Since velocities are sourced by growth of overdensities, the cosmological density and velocity fields are correlated in the linear regime;
however, one can ignore such correlations at arcminute scales  where non-linear evolution of matter takes over \citep{H2000}. The coherence length of velocity fluctuations are of the order $\sim$ 50 Mpc (comoving), therefore, we can assume velocity to be constant on smaller length scales.

The anisotropies or fluctuations due to \ksz secondary CMB distortion is proportional to fluctuations of number density electrons. The local electron number fluctuation can broadly be categorized into two varieties. 
First, when the universe is completely ionized, the fluctuations in kSZ are proportional to the density modulation of electrons. These hot electrons are typically inside
dark matter halos and can be computed analytically using halo model \citep{S2000,H2000, MF2002}. Secondly, electron number fluctuations can also occur due to ionization fraction modulation.  Several numerical simulations \citep{MFHZZ2005,IPBMS2007,MMS2012,SRN2012,BTCL2013,A2016} have shown that   both the amplitude and the shape of kSZ anisotropies is sensitive to the details of the epoch of reionization. For example, nascent structures in the initially neutral Universe can emit ionizing photons giving rise to isolated ionized bubbles and, hence, ionization fraction modulated regions.

The signature of the whole reionization process is imprinted on the \ksz distortions. As long as the ionized bubbles keep their individual existence, the  
 multipole at which kSZ power spectrum reaches its maximum is given by the `characteristic' size of the bubbles \citep{GH1998}. As reionization proceeds, these bubbles start to overlap and, the ionization fluctuations reduce as late time density fluctuations due to formation of halos, take over. The late kSZ (due to density fluctuations) and kSZ from patchy reionization contribute roughly equal to the total kSZ from reionization \citep{SBDHHLOT2010,websky2020} but patchy-kSZ is sensitive to the details of structure formation. Recently, there has been claims of detection of patchy-kSZ at $\ell=3000$ \citep{SPTksz2021} using simple parametric fits to simulations \citep{BNTCL2013} (but also see concerns raised in  \cite{PMC2021} and \cite{GIDAL2020}).

A precise measurement of patchy-kSZ has significant implications for the understanding of reionization since the source which dominates patchy-kSZ is also expected to drive reionization. The preferred candidate for causing reionization are early galaxies hosted within low-mass haloes ($\gtrsim 10^9-10^{10}M_{\odot}$) which are abundant at $z\gtrsim 6$. Another possible source for reionization can be high redshift quasars.  This patchy-kSZ from quasars, assuming them to be the dominant source of reionization, was first computed in \cite{ADPG1996}, and it opened up the field of SZ studies from sources other than galaxy clusters. However, one needs to do an improved calculation with more recent data available today \citep{Pl2016tSZ,ACT2020,ACTtsz2020,SPTksz2021}. Quasars being more luminous than galaxies, the ionized bubbles around quasars are bigger and can reach $\sim$ 10-20 Mpc in size \citep{Fan2006,BH2007}. The size of the bubbles are governed by the speed at which the region around a quasar is ionized and is rapid due to the  photon propagation speed. In contrast, the central quasar moves a tiny distance during the lifetime of the bubble, and the bubble can be thought to be intact and ionized around the quasar. The medium inside the bubble flows with the local bulk velocity before and after ionization, and hence the ionized bubble acts as new local source of \ksz. This opens up the exciting prospect of detecting individual quasars bubbles in the CMB sky which can used as probe of both the IGM at high redshifts and quasar astrophysics. 

In this work, we compute the patchy-kSZ contribution from quasars, improving upon the work of \cite{ADPG1996}, using recent knowledge of quasar abundance and luminosity properties \citep{SHFARRH2020}. We first perform simple analytic estimate of the size of ionized bubbles around the quasars and follow it up with a more rigorous 1-dimensional, spherically symmetric, radiative transfer code calculation \citep{BH2007,CG2021}, in Sec. \ref{sec:ionized_bubble}. The results from the radiative transfer codes, developed for the computation of quasar proximity zones, agree with the analytic estimates of the size of ionized bubbles as shown in Appendix A of \cite{CG2021}. Then, we briefly overview the Quasar luminosity function of \cite{SHFARRH2020}. Using the bubble size calculations, we set up a halo model prescription for quasars \citep{MW2001,HH2001}. We extrapolate this halo model using the observed quasar abundance to higher redshifts and compute kSZ power spectrum during reionization from quasars. In these calculations, we assume the ionized bubbles to be well separated which is a good approximation till the time when the bubbles start to overlap. 

Once the bubbles start to overlap, the anisotropies are expected to reduce. Therefore, our computation with separated bubbles can be seen as an upper limit to the kSZ distortion from quasars. The final \ksz power spectrum from quasars is found to be subdominant compared to reionization-kSZ obtained from numerical simulations \citep{SBDHHLOT2010} which we will show in Sec. \ref{sec:kSZ_quasar}. Finally, we discuss the prospects of detecting individual quasar bubbles in kSZ observations using matched filtering \citep{HT1996,MBD2006}. As the spectrum of kSZ is a black body, we can not differentiate kSZ from primary CMB using only spectral information. Hence, we need to use spatial anisotropy information of primary CMB and the quasar bubble kSZ to distinguish the two. We show that with futuristic high resolution, higher sensitivity ,  experiments such as \cite{CMB_HD}, we may have a reasonable chance of detecting individual quasar bubbles given we are able to remove the foregrounds.   

\section{Ionization bubble from quasar}
\subsection{Growth of the Str$\ddot{o}$mgren sphere}
\label{sec:ionized_bubble}
Quasars are sources of energetic photons which can ionize and heat the surrounding gas before reionization ($z\gtrsim 6$). The spectrum is given by, $L_{\nu}=L_0\left(\frac{\nu}{13.6}\right)^{-\alpha}$,
where $L_\nu$ is the luminosity. The number of photons emitted per unit time is connected to the luminosity of a quasar and given by, $\dot{N}=\int_{13.6} ^\infty \left(\frac{L_{\nu}}{h\nu}\right)d\nu$. For the sake of simplicity, we consider the medium surrounding quasars to only consist of hydrogen. To compute the size of an ionized bubble, at first approximation, we assume each radiated photon to ionize a single neutral hydrogen atom. The medium close to the quasar will be significantly ionized as photons outnumber the local density of hydrogen atom significantly. However, as the distance from the quasar increases, the number of available photons for ionizing is reduced as they have already been consumed near the quasar and,furthermore, the number of hydrogen atoms increases being proportional to square of the distance. Therefore, at some distance from the quasar, the available photons are fully spent which determines the characteristic size of the ionization bubble. The evolution of ionization front, assuming spherical symmetry, is determined by the balance of ionization-vs-recombination \citep{SIAS2006,BH2007},
\begin{equation}
    4\pi R^2f_Hn_H\frac{dR}{dt}=\dot{N}-\frac{4}{3}\pi R^3\alpha_{HII}n^2_H,
    \label{eq:ionization_front}
\end{equation}
where $R$ is the size of Str$\ddot{o}$mgren sphere, $f_H$ is the fraction of neutral hydrogen having number density $n_H$, and $\alpha_{HII}(T)$ is the recombination coefficient which is a function of the temperature. The recombination coefficient is of the order of $10^{-13} {\rm cm}^3$s$^{-1}$. Fitting formulae (used in our Appendix \ref{app:rates}) to obtain the heating and cooling rates are taken from Appendix B of \cite{BH2007}.

Note, that a photon with energy ($E$) greater than 13.6 eV will impart the excess energy to heat the ambient medium. 
 The average energy imparted to the electrons by the ionizing photons is given by $<E_e>=\frac{\int \left(L_{\nu}/h\nu\right)(h\nu-13.6)\sigma(\nu)d\nu}{\int\left(L_{\nu}/h\nu\right)\sigma(\nu)d\nu}\,$=\,3.33 eV where $\sigma(\nu)$ is the photo-ionization cross-section (Eq. \ref{eq:photoionization}) for neutral hydrogen and has a steeply decreasing power law relation with increasing photon energy and a fiducial $\alpha=1.5$. This average energy corresponds to the  temperature of the medium of $\approx 2\times10^4$ K.


\begin{figure}
\centering 
\includegraphics[width=\columnwidth]{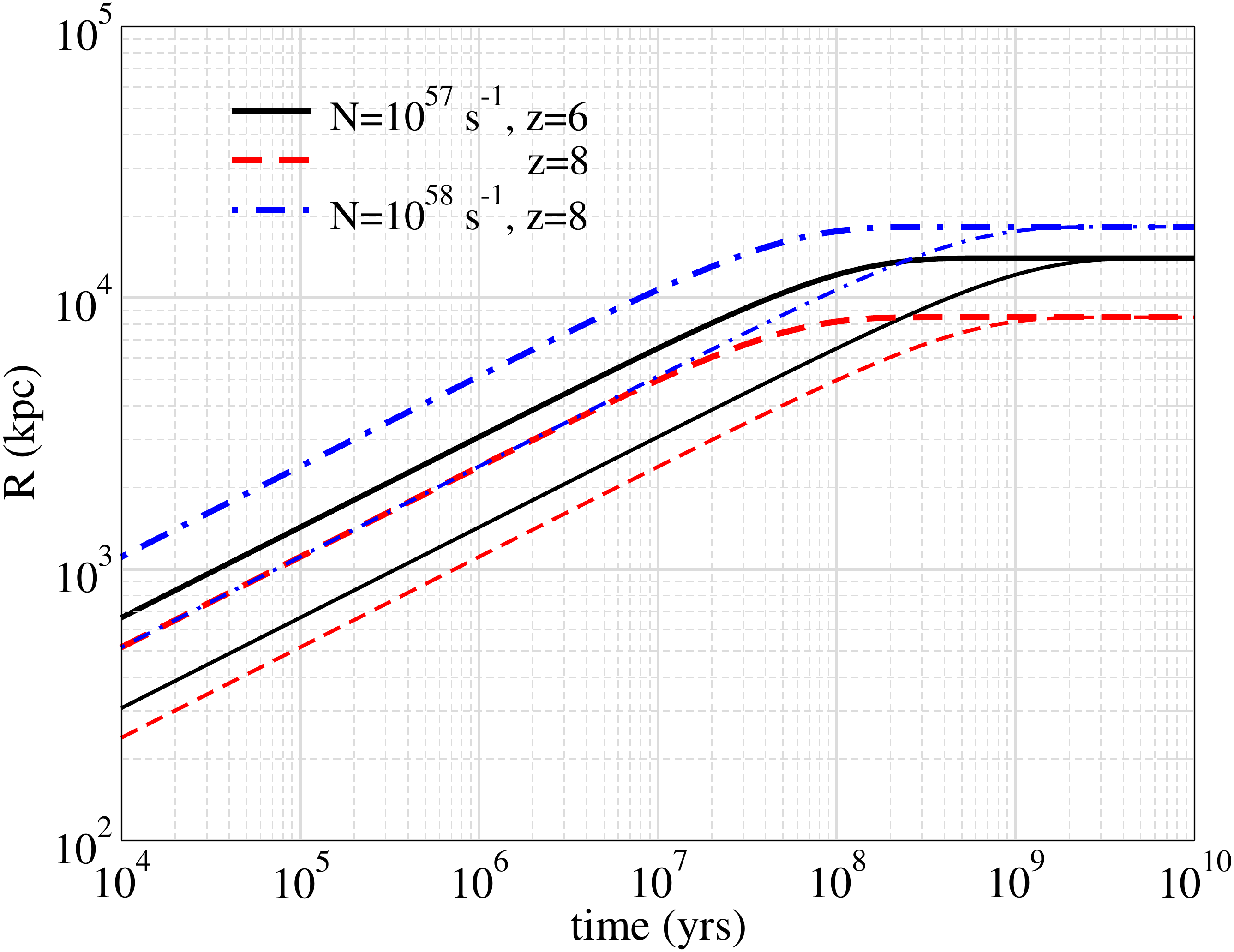}
\caption{Evolution of the ionization front as a function of time around a quasar at $z=6$, $\dot{N}=10^{57}$s$^{-1}$ (black solid), $z=8$, $\dot{N}=10^{57}$s$^{-1}$ (red dashed) and $z=8$, $\dot{N}=10^{58}$s$^{-1}$ (blue dash-dotted) for two different neutral hydrogen density $f_H=0.1$ (thick lines) and $f_H=1$  (thin lines). 
}  
\label{fig:ionization_front}
\end{figure}

The solution to Eq. \ref{eq:ionization_front} gives the time evolution of the size of the ionized bubble,
\begin{equation}
    R(t)= R_S \left[1-\rm{exp}\left(-t / f_Ht_{rec}\right)\right]^{1/3},
    \label{eq:Ifront_analytic}
\end{equation}
where $R_S=\left(\frac{3\dot{N}}{4\pi\alpha_{HI}n^2_H}\right)^{1/3}$ is the so called Str$\ddot{o}$mgren radius, $t_{rec}=1/(\alpha_{HII}n_{H})$ and we have taken $n_H$ to be the mean cosmological density at a given redshift.  Fig. \ref{eq:ionization_front} shows the time evolution of the ionized bubble size for the fully neutral medium (i.e $\rm{f_H} = 1$) and a fractionally ionized medium at the redshift $z=6$. Note, that for a partially ionized medium, the size of the ionization front is larger compared to the same in neutral medium at a given time, the reason being that there are less number of neutral hydrogen to ionize and the front can advance until recombination becomes effective. 
For the hydrogen density at $z=6$ and $T\sim 10^4$K, the recombination time scale is of the order of $10^9$ yr. Hence, once, recombination becomes effective, the ionization front does not grow in size anymore as each ionization is nullified by a recombination. 

Apart from the simple estimate shown above, one can do more involved 1-D radiative transfer calculations to obtain the ionization profile around a quasar \citep{BH2007,CG2021,SKKH2022}. These calculation have been done to obtain a measure quasar proximity zones which 
is defined as the distances from individual quasar where the transmission probability of Lyman-alpha flux reduces to 10 percent \citep{BH2007} i.e., $e^{-\tau}=0.1$, where $\tau$ is the optical depth of Lyman-alpha scattering for neutral hydrogen. This optical depth $\tau$ becomes comparable to 1 even for neutral hydrogen fraction as low as $\sim 10^{-4}$. Even when the medium is practically ionized, a tiny amount of neutral hydrogen can trap the Lyman-alpha radiation due to the high cross-section. Therefore, for calculations of quasar proximity zones, one needs to calculate the density profile of neutral hydrogen accurately and such 1-D radiative transfer calculations are necessary. For the \ksz signal that we are interested in this work, the ionization front radius is the relevant quantity to compute. \cite{CG2021} have shown that Eq. \ref{eq:Ifront_analytic} matches very well with profiles obtained from detailed radiative calculations (see the  middle panel of their Fig. 15). 
In the current work, we use the analytic estimates  of the growth of ionized region throughout, since the time consuming full radiative transfer calculations give very similar results.  These calculations also give temperature of the bubble medium to be $\sim 10^4$ K.  One needs to do full hydrodynamical  cosmological simulations to understand the density inhomogeneities inside the ionized bubbles; however, in this work,  we use homogenous cosmological density inside the bubbles.

\subsection{1D radiative transfer problem for evolution of the ionization front}
\label{subsec:radiative_transfer}
The evolution of ionization fraction and temperature is given by \cite{C1992,HG1997,BH2007,TZ2008},
\begin{equation}
    \frac{{\rm d}x_e}{{\rm d}t}=n_{HI}(\Gamma^{\gamma}+n_e\Gamma^{e})-n_en_{HI}\alpha_{HII}
    \label{eq:ionization_fraction}
\end{equation}
\begin{equation}
    \frac{{\rm d}T}{{\rm d}t}=\frac{2}{3k_B}(H_{tot}-\Lambda_{tot})-2HT
      \label{eq:temperature_eq}
\end{equation}
In Eq. \ref{eq:ionization_fraction}, $\Gamma^{\gamma}$ is the photo-ionization rate and $\Gamma^{e}$ is the ionization rate from the secondary electrons. We  
ignore the secondary ionization term for our calculation since including it makes little difference \citep{CG2021} due to the steep photon spectrum which ensures that most of the ionizing photons have energy comparable to ionization threshold.  
The coefficients for photo-ionization, recombination, heating and cooling can be found in \cite{C1992,BH2007}. In the spherical symmetry approximation, we divide the medium surrounding the quasar into concentric spherical cells. We then evolve each cell for a time step starting from the smallest cells and then move outwards. In each cell, some photons are absorbed and the surviving photons move on to next cell. The incident photon radiation flux with energy $E(h\nu)$ on a given a cell `i' is given by,
\begin{equation}
    L'_i(\nu,t)=e^{-\tau_i(\nu,t)}L_{\nu},
\end{equation}
 where $L_{\nu}$ is the luminosity at the zeroth cell and the optical depth $\tau_i(\nu,t)=\sum_{j}\sigma_{HI}(\nu)n_{HI,j}(t)dr$, where $j$ is the index of all cells preceding the cell $i$ i.e $j<i$.  Note, that we have to take into account the number density of neutral hydrogen in a given cell. With this setup, we solve for all spatial cells for a given timestep and then move onto the next timestep.


\begin{figure}
\centering 
\includegraphics[width=\columnwidth]{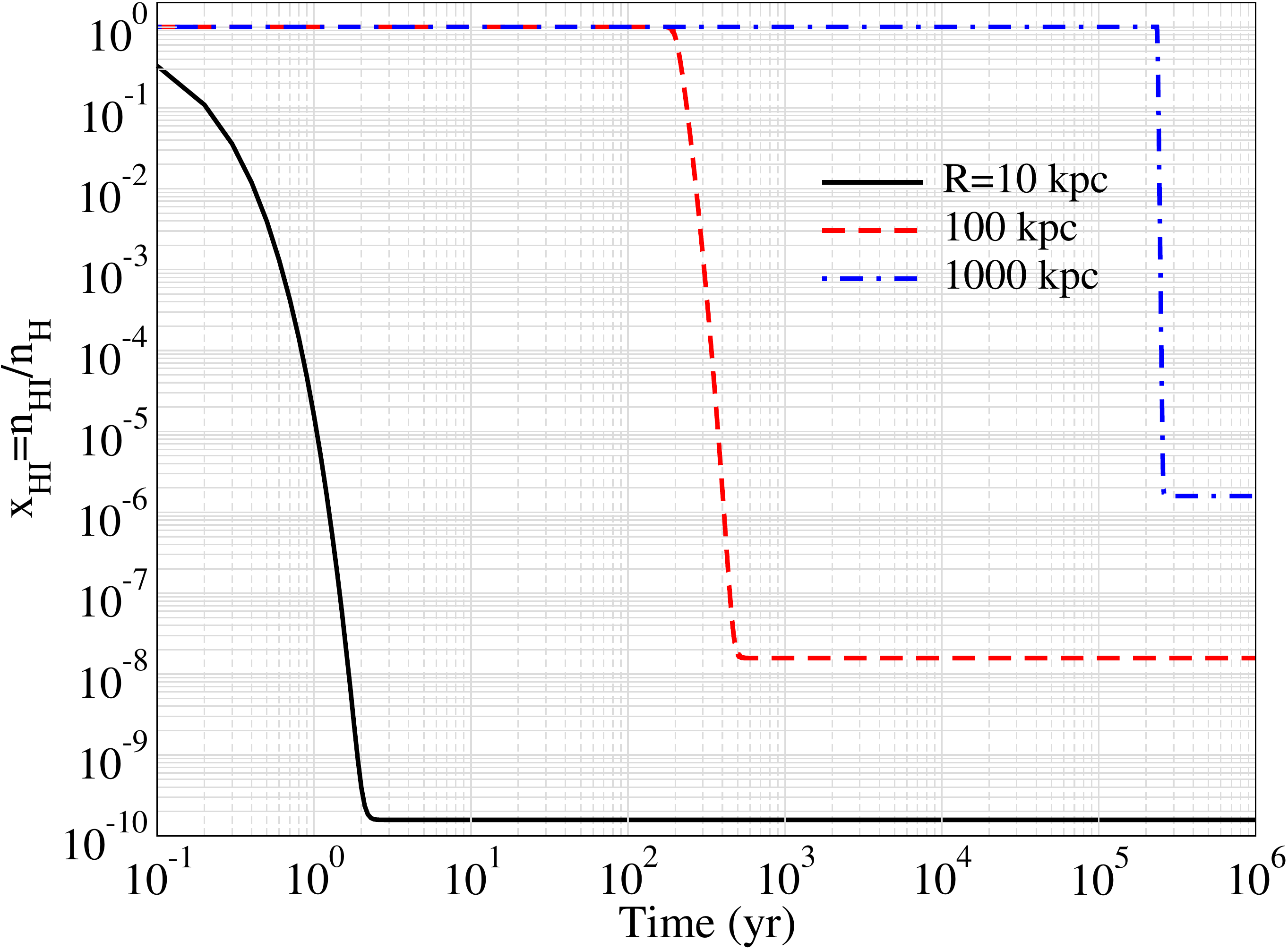}
\caption{Evolution of neutral hydrogen fraction for a few spatial cells which are at a distance of 10, 100 and 1000 kpc away from the quasar. }  
\label{fig:1st_shell_evolution}
\end{figure}

\begin{figure*}
\centering 
\includegraphics[width=\columnwidth]{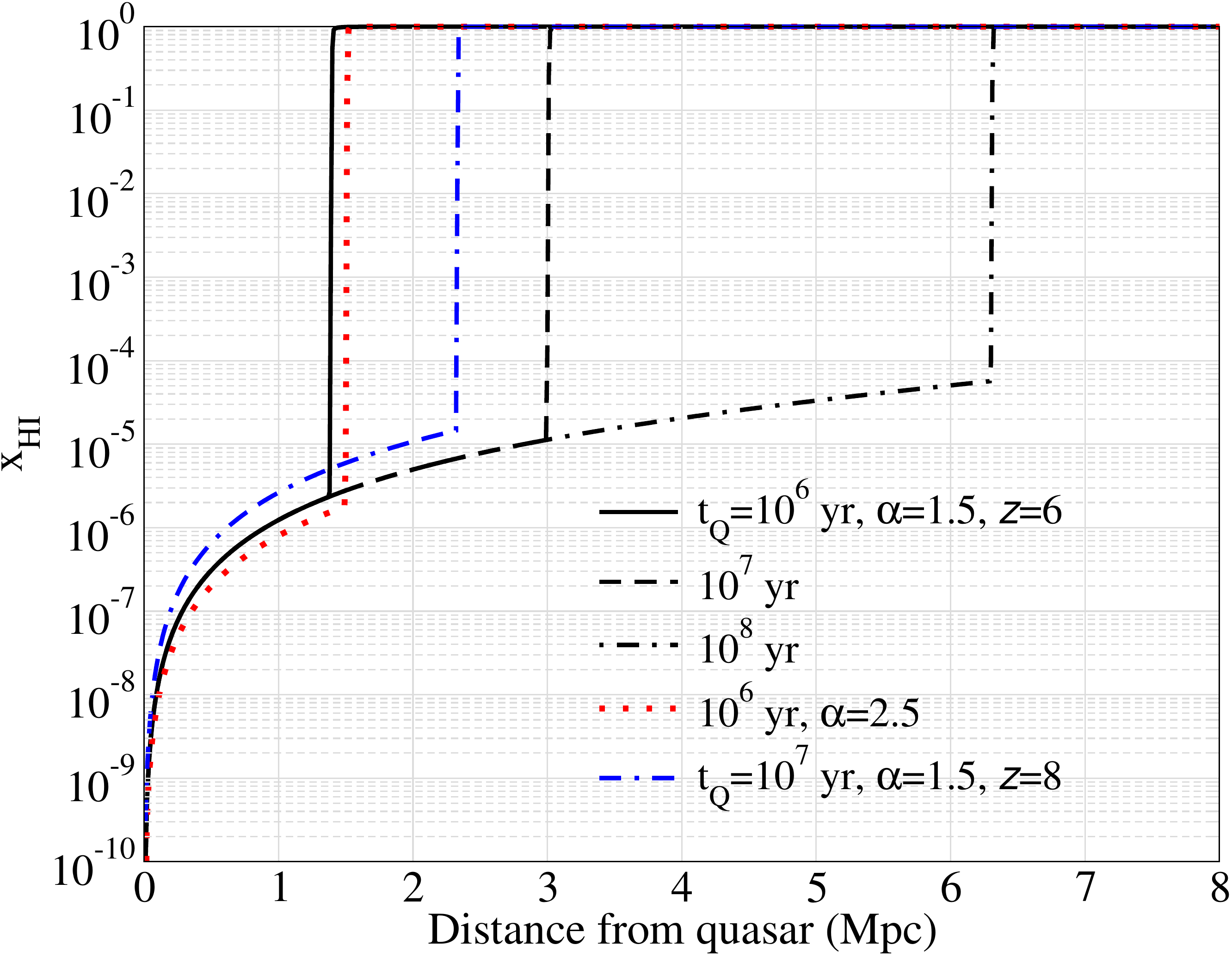}
\hspace{4mm}
\includegraphics[width=\columnwidth]{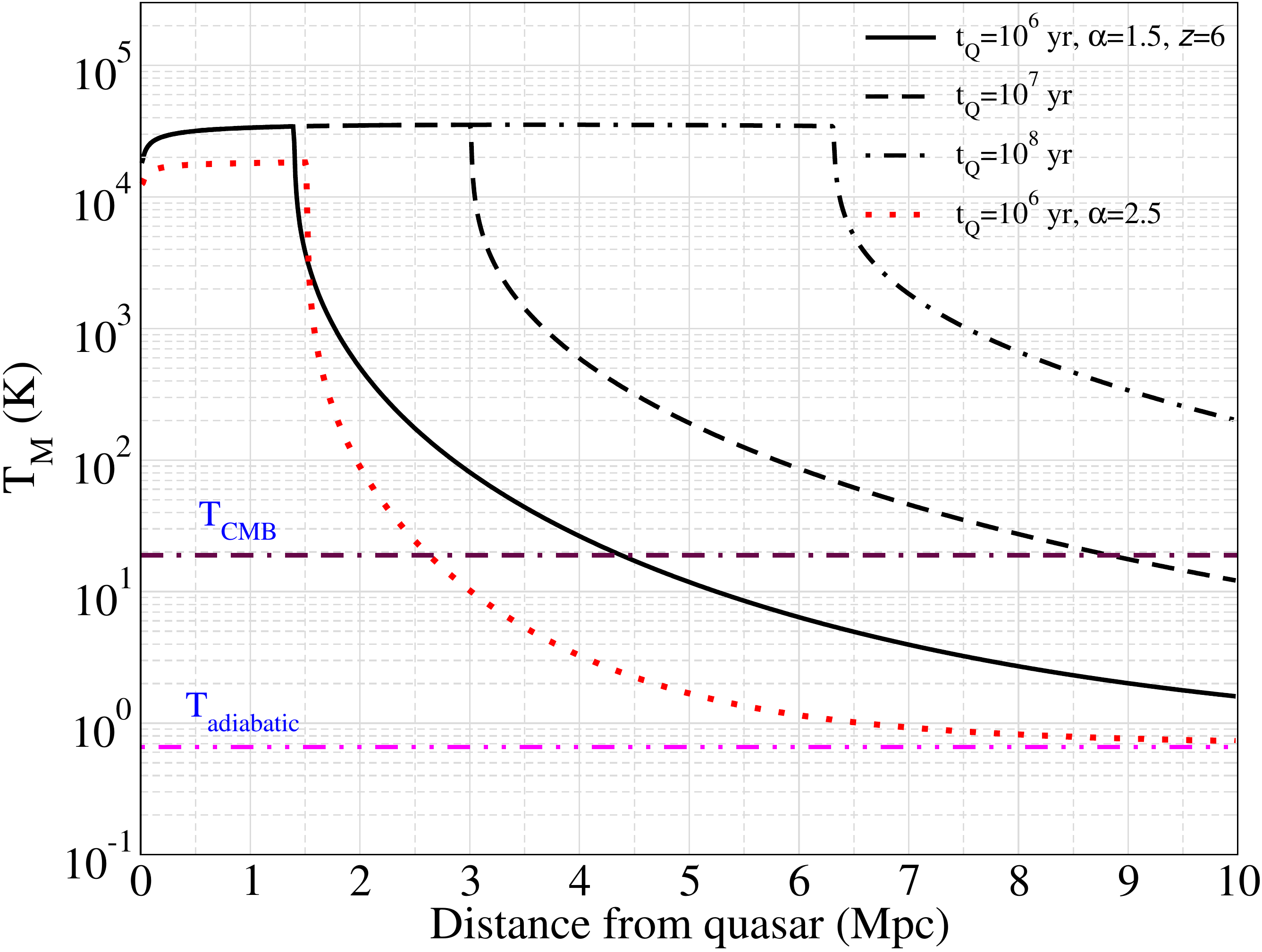}
\caption{Profile of (a)neutral hydrogen fraction, (b)gas temperature as a function of distance from the quasar with  $\dot{N}=10^{57}$s$^{-1}$, redshift as shown in the figure and assuming that the quasar is surrounded by neutral hydrogen. We also show CMB and adiabatic matter temperature in the right panel. 
}  
\label{fig:xHI_evolution}
\end{figure*}


 We show the temperature dependence of the cooling rates for different physical processes in Fig. \ref{fig:cooling_rates}. It is evident that photo-heating is the dominant process at least close to the quasar. In Fig. \ref{fig:1st_shell_evolution}, we show the time evolution of the ionization fraction of the first cell at a distances of 10, 100 \& 1000 kpc from quasar. As we move outward from the quasar, this timescale increases as the ionization front moves slowly outwards. Once the number density of neutral hydrogen  becomes sufficiently small, at a given distance from the quasar, the photo-ionization rate and recombination rate match each other and a local equilibrium in the ionization fraction is established. It is evident from the figure that ionized bubble grows from $\sim 1$ kpc to $\sim 1$ Mpc within timescales of $1-10^5$ years. Thus, multiple timescales are involved in the evolution of the quasar bubble\footnote{ We incorporate a global timestep whose value is constrained by the first cell}.  Note, that these timescales also mean that large bubble are quickly ionized in time period far shorter than the lifetime of an active quasar or the Hubble time. These bubbles remain as active sources of kSZ distortions throughout the lifetime of each quasar. Once the quasars die, `dead bubbles' may still exist and continue to distort the CMB.

We plot the neutral hydrogen fraction and the temperature profiles, as a function of the distance from the quasar, in Fig. \ref{fig:xHI_evolution}. It has been shown that the size of ionization bubble from the detailed radiative transfer calculation matches very well with the simplistic calculation of Fig. \ref{fig:ionization_front} (Appendix A of \cite{CG2021}). From Fig. \ref{fig:xHI_evolution}, we can see that the temperature falls off much more smoothly beyond the ionization front. Outside the ionized bubble, recombination catches up with photoionization leading to no change in ionization fraction. However, each ionization creates a free electron with some kinetic energy which is shared with the ambient medium which causes the temperature to decrease smoothly rather than abruptly. Also, note that different spectral indices leads to significant differences in temperature profile. This is related to the fact that the more energetic photons in the power law tail can survive to longer distances from the quasar as the photo-ionization cross-section falls as a steep power law. In fact, for a  steeper photon spectrum, more number of photons are on the ionization threshold which get trapped closer to the quasar.

\section{Number density of quasar bubbles}
\label{sec:bubble_distribution}
To compute the kSZ power spectrum from quasars, we need to know the number density of quasars as a function of redshift. The comoving number density of quasars as a function of luminosity, at a particular redshift, is given by the quasar luminosity function. We begin by describing the quasar luminosity function  following the recent results given by \cite{SHFARRH2020}. Next, we use a halo model formalism to compute the number density of quasars following \cite{HH2001} and \cite{MW2001}. 

\subsection{Quasar luminosity function approach}
\label{subsec:shen_2020_results}
Following \citep{SHFARRH2020}, the quasar luminosity function (QLF), upto $z\sim 7$, is well described by the form
\begin{equation}
    \phi(L_{\alpha},z)=\frac{\Phi^*}{L^*_{\alpha}(z)}\left[\left(\frac{L_{\alpha}}{L^*_{\alpha}(z)}\right)^{\beta_1}+\left(\frac{L_{\alpha}}{L^*_{\alpha}(z)}\right)^{\beta_2}\right]^{-1}\; ,
    \label{eq:quasar_luminosity_function}
\end{equation}
where the fitting functions are given by
\begin{equation}
\beta_1(z)=1+a_0T_0(1+z)+a_1T_1(1+z)+a_2T_2(1+z)\nonumber
\end{equation}
\begin{equation}
\beta_2(z)=1+\frac{2b_0}{\left(\frac{1+z}{1+z_{\rm ref}}\right)^{b_1}+\left(\frac{1+z}{1+z_{\rm ref}}\right)^{b_2}}\nonumber
\end{equation}
\begin{equation}
{\rm log}L_{\alpha}^*(z)=\frac{2c_0}{\left(\frac{1+z}{1+z_{\rm ref}}\right)^{c_1}+\left(\frac{1+z}{1+z_{\rm ref}}\right)^{c_2}}\nonumber
\end{equation}
\begin{equation}
\Phi^*(z)=\frac{d_0T_0(1+z)+d_1T_1(1+z)+d_2T_2(1+z)}{\rm ln10},
\end{equation}
where $T_0(x)=1$, $T_1(x)=x$ and $T_2(x)=2x^2-1$.
We use the fitting parameters for Global fit A model which is given in Table 3 of \cite{SHFARRH2020}. The number density of quasars at a given redshift can then be written as
\begin{equation}
n_Q(z)=\int {\rm d}L_{\alpha}\phi(L_{\alpha},z)
\end{equation}
At $z\sim 6$, the detected quasars have bolometric luminosity $L_{\rm bol}>10^{45}$ erg$s^{-1}$ (Fig. 5 of \cite{SHFARRH2020}). We, therefore, choose the integral limit in the above integral between $L_{\alpha}=10^{45}-10^{47}$ ergs$^{-1}$. We note that the characteristic luminosity at $z\sim 6$ is of the order of $\approx 10^{46}$ ergs$^{-1}$ beyond which the quasar luminosity function has a steep decline. This gives us a conservative estimate on the number density of quasars. Also, note that we extrapolate the fitting formulae to obtain the quasar number densities at $z > 7$. The luminosity in the UV band is related to the bolometric luminosity with log($L_{\rm bol}/L_{\rm UV})\approx$ 0.8 (Fig. 2 of \cite{SHFARRH2020}). Assuming each photon in UV band leads to one ionization of neutral hydrogen, the rate of production of ionizing photons for the characteristic luminosity leads to our fiducial $\dot{N}\approx 10^{56}$ s$^{-1}$ at $z\sim 6$.   

\subsection{Halo model for quasar bubble distribution}
\label{subsec:halo_model}
An alternate route based on the halo model to estimate the space density of quasars was taken by \cite{HH2001} \& \cite{MW2001}. In this approach, the abundance of quasars is given by
\begin{equation}
    \Phi(z)=\int_{M_{min}} ^{\infty}\frac{t_Q}{t_H}n(M){\rm d}M,
    \label{eq:halo_quasar}
\end{equation}
where $t_Q, t_H \,\& \, n(M)$ are the quasar lifetime, the halo survival time and the halo mass function, respectively. The expression for halo mass function is given by,
\begin{equation}
\frac{{\rm d}n}{{\rm d}M}=f(\sigma,z)\frac{\rho_{m0}}{M}\frac{{\rm d}ln\sigma^{-1}}{{\rm d}M},
\label{eq:massfunction}
\end{equation} 
We use the fitting function in \cite{TKKAWYGH2008, TRKKWYG2010} for $f(\sigma,z)$ given by
   \begin{equation}
   f(\sigma,z)=A\left[\left({\frac{\sigma}{b}}\right)^{-a}+1\right]e^{-(c/\sigma^2)}
   \end{equation}
   with $A=0.26,a=2.3,b=1.46,c=1.97$.
 We define the halo mass using the over-density of 1600 w.r.t the mean matter density. 

In this halo model approach to quasar abundance, each halo is assumed to have at most one quasar with the probability given by the duty cycle, $f_Q=\frac{t_Q}{t_H}$, where the  halo survival time is the timescale over which halos merge to produce another halo of a higher mass. This timescale is close to the Hubble time \citep{MW2001} and, hence, we use the Hubble time as $t_H$. Thus, $f_Q$ is the probability of a quasar, with a given lifetime, to be active during the timescale over which halos of a given mass survive. The number density of quasars, $n_Q(z)$, is then given by,
\begin{equation}
n_Q(z) = \int_{M_{min}} ^{\infty}f_Q n(M)dM,
\label{eq:number_density}
\end{equation}
where $M_{min}$ is the minimum halo mass required to host a quasar.
 From the above equation, it is clear that there exists a degeneracy between the duty cycle and the limiting mass. \cite{S2007} used the quasar luminosity function up to $z\approx 5$ to constrain $M_{min}$ for a given duty cycle
 
This degeneracy can be broken by measuring quasar clustering\footnote{This is in the same spirit of self-calibration in galaxy cluster surveys  \citep{MM2004}}.  Clustering gives a measure of mass-dependent bias, $b_{\rm eff}(M_{\rm h})$), given by
\begin{equation}
    b_{\rm eff}(M_{\rm h})=\frac{\int_{M_h} ^{\infty}b(M)n(M)dM}{\int_{M_h} ^{\infty}n(M)dM}
    \label{eq:bias}
\end{equation}
To a good approximation, $b_{\rm eff}(M_{\rm h})$ can be approximated to be $b_{\rm eff}(M_{\rm min})$ as the halo mass function is a steeply decreasing function of the halo mass.
Quasar clustering measurements \citep{S2007,E2015} have been used to constrain 
 $M_{min} \approx 10^{12} M_{\odot}$. Using this information, the quasar lifetime can be constrained to be $\approx 10^6-10^7$ yr for $z<3.5$ while for $z>3.5$, $t_Q$ is of the order $\gtrsim 10^8$ yr \citep{S2007} (see also Table 7 in \cite{E2015}). In our calculations for SZ fluctuations from quasar bubbles, we use $10^7$ yr as the fiducial quasar lifetime. For a fixed `observed' number density of quasars, as we increase $t_Q$, the increase in the quasar number is balanced by increasing $M_{min}$ which leads to an decrease in the number of quasar host halos.  
 
\begin{figure}
\centering 
\includegraphics[width=\columnwidth]{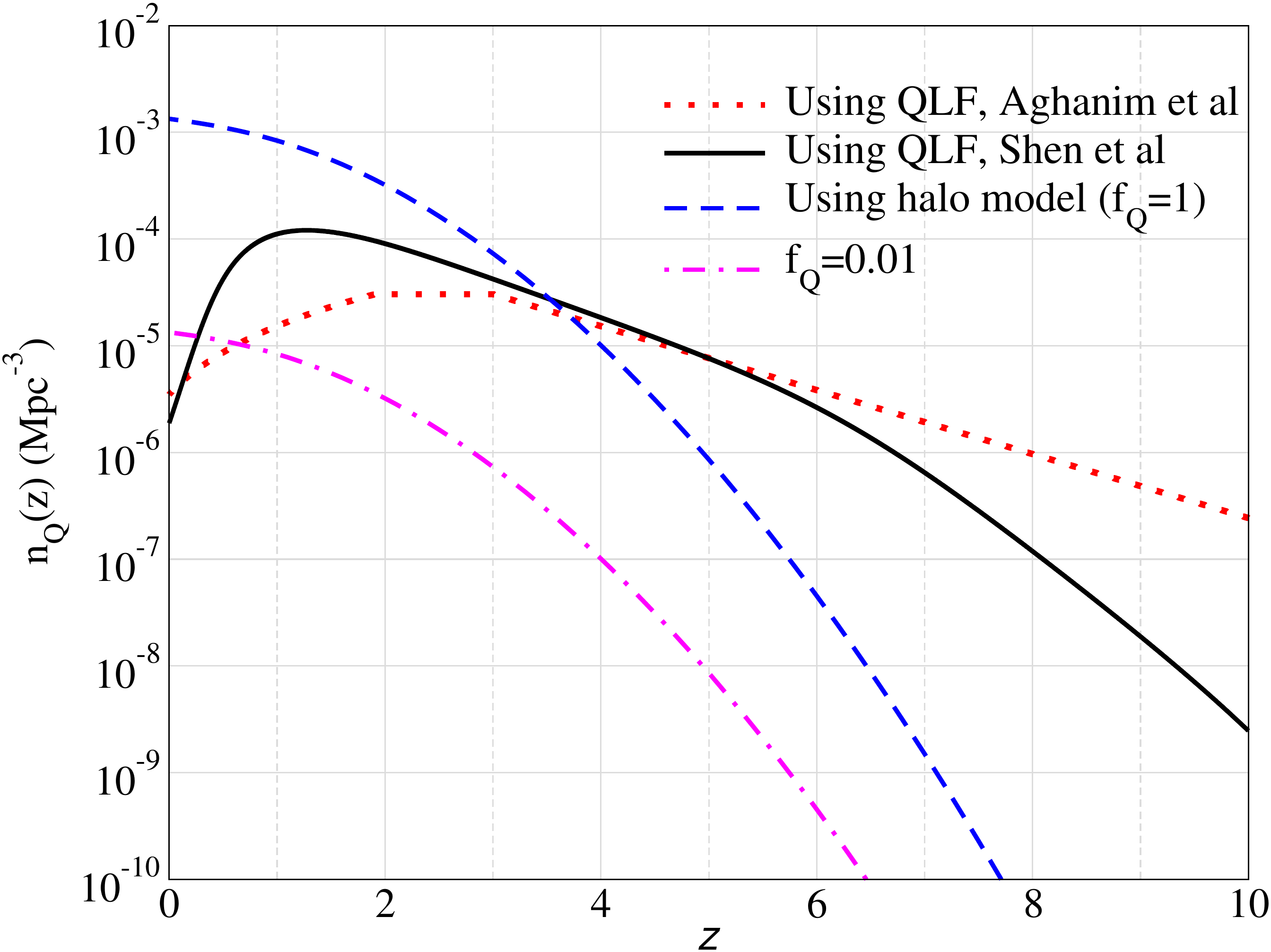}
\caption{Quasar abundance predicted using QLF by \citet{SHFARRH2020}, and from halo model with duty cycles $f_Q$ =1 \& 0.01. The halo model calculations show the number of halos above the threshold $10^{12} M_{\odot}$ for both $f_Q$.   For comparison, we also show the quasar abundance used in \citet{ADPG1996}.  } 
\label{fig:aghanim_vs_halo_model}
\end{figure}

 In Fig. \ref{fig:aghanim_vs_halo_model}, we show the comparison between the quasar abundance predicted using the QLF of \cite{SHFARRH2020} versus the same from the halo model (See also \citep{RT2021}). We note that at $z>4$, halo model severely underpredicts the number of quasars. This would mean that $M_{min}$ has to be reduced in order to accommodate the number density of quasars given $f_Q=1$. We would remind the reader that we have extrapolated the QLF as well as the halo model to higher redshifts, well beyond they are currently constrained by the data. \\

\section{Observables from quasar bubbles}
\label{sec:quasar_observable}

\begin{figure}
\centering 
\includegraphics[width=\columnwidth]{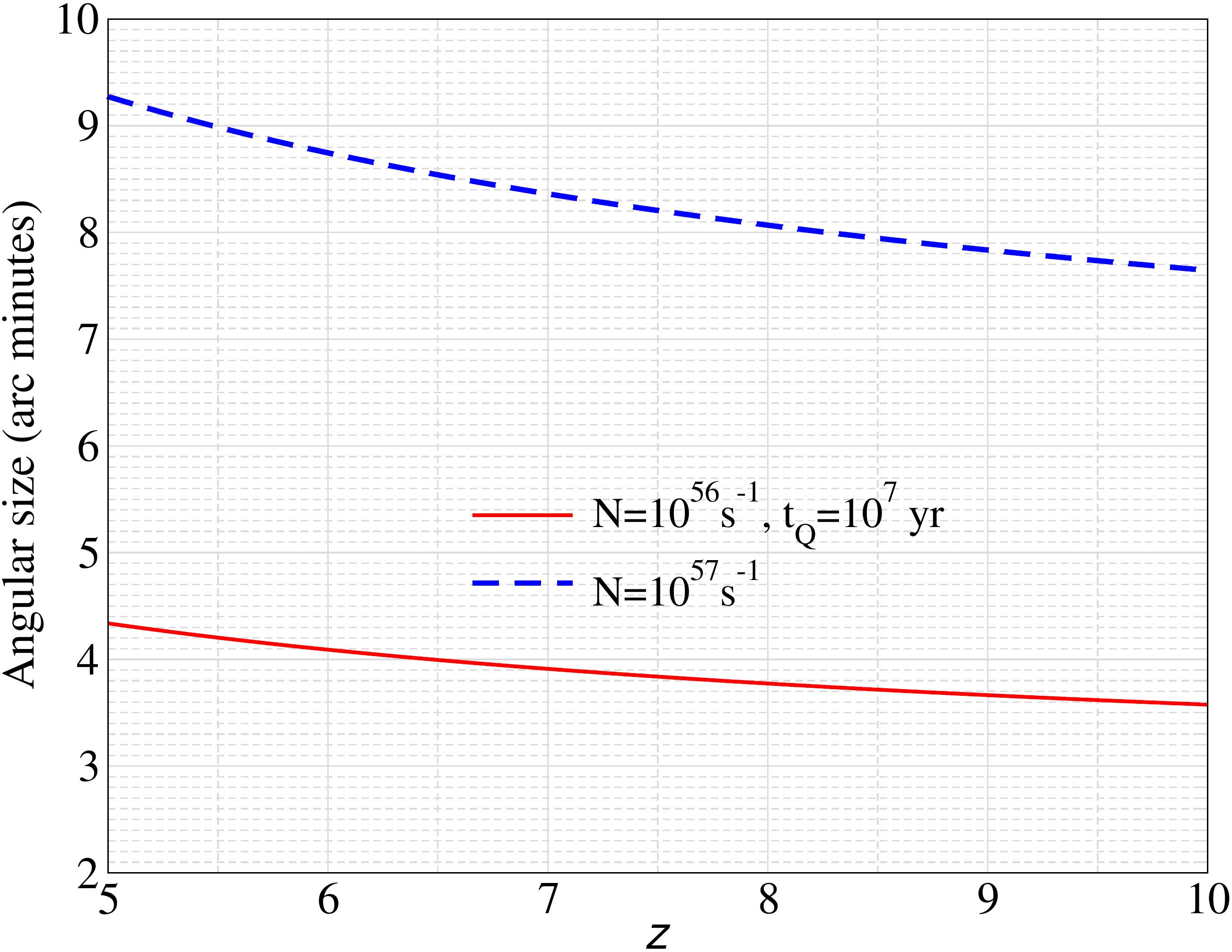}
\includegraphics[width=\columnwidth]{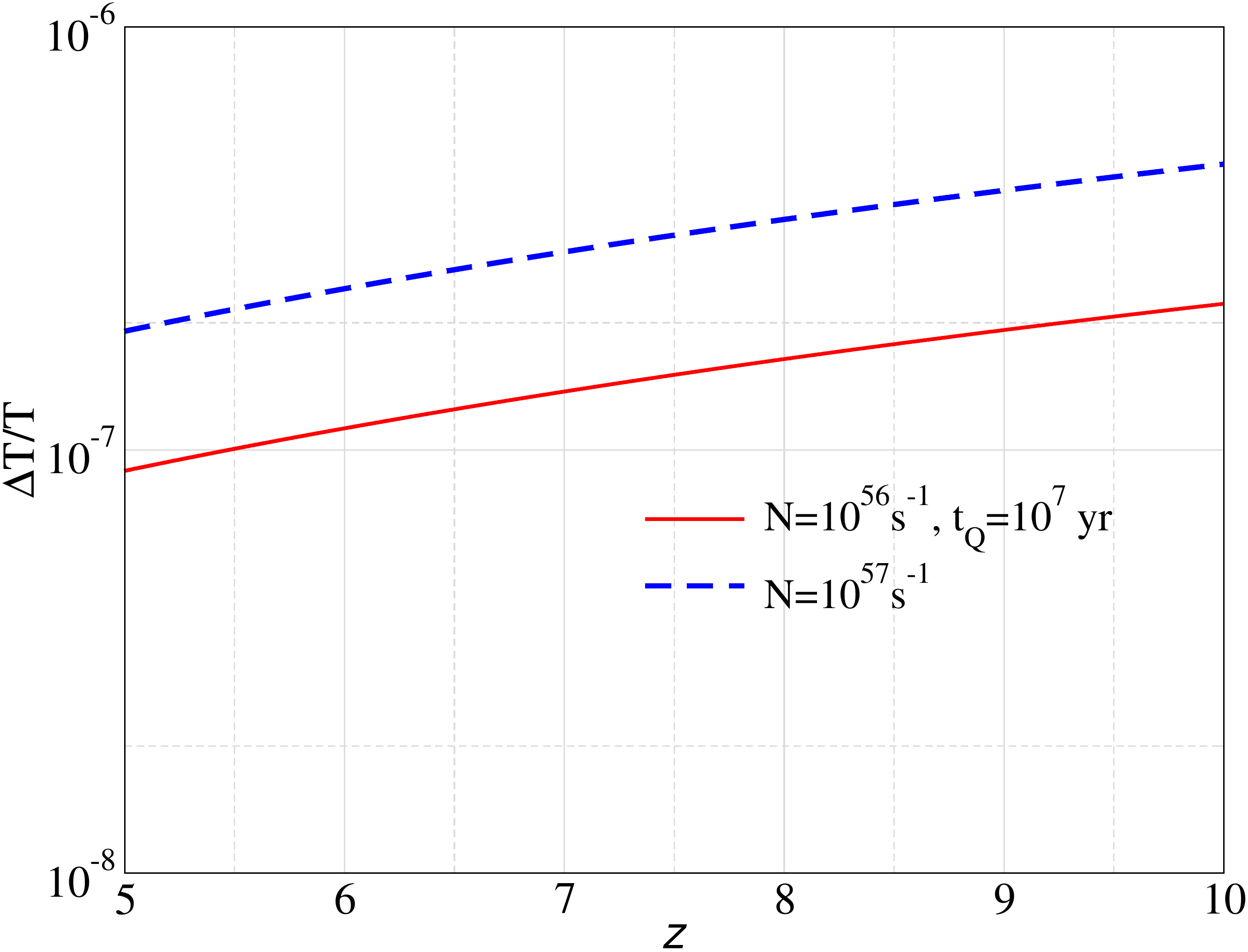}
\caption{Angular sizes and the kSZ decrements along the line of sight through the centre of quasar bubbles as a function of redshift. 
}
\label{fig:quasar_optical_depth}
\end{figure}
Once the number densities of the quasar bubbles were estimated, as discussed above, we can use the analytic formula given Eq. in \ref{eq:Ifront_analytic} for  estimating the size of the bubbles. For quasar lifetimes $t_Q<< t_{rec}$, we have the size of the bubble  given by $R(z)=\left(\frac{3\dot{N}t_Q}{4\pi n_{H(z=0)}}\right)^{1/3}(1+z)^{-1}$ (Eq. \ref{eq:Ifront_analytic}). 
This gives a typical size $R(z)\approx 10\,(1+z)^{-1}$ Mpc for 
 a characteristic luminosity of the order $10^{46}$ergs$^{-1}$ which corresponds to $\dot{N}\approx 10^{56}$ s$^{-1}$ and $t_Q=10^7$yr. In Fig. \ref{fig:quasar_optical_depth}, we plot the angular size subtended by the ionization bubble as a function of redshift. 
 The typical angular scale subtended by  such a bubble lies roughly between $4-10$ arc-mins for $z > 5$. The corresponding  optical depth  is given by $\tau=2\sigma_{\rm T} n_e R$ where $\sigma_{\rm T}$ is the Thomson cross-section and $n_e$ is the background electron density at redshift $z$.
 
The ionized electrons inside the bubble will give rise to both tSZ and kSZ signal. The amplitude of kSZ signal is given by, $\frac{\delta T}{T}=\frac{v}{c}\tau$ \citep{SZ1980}, where $v/c$ is of the order $10^{-3}$, whereas for tSZ, $\frac{\delta T}{T}=\frac{T_e}{m_e}\tau$ \citep{ZS1969}. As shown in Fig. \ref{fig:xHI_evolution}, the gas temperature inside the bubble is $\sim 10^4$ K (i.e.,$\sim$ 1 eV) and, hence, the \ksz dominates the \tsz by 2-3 orders of magnitude. 
We show the the redshift dependence of the bubble angular size and temperature fluctuation in Fig. \ref{fig:quasar_optical_depth}. For higher $\dot{N}$, the size of the bubble is larger, which results in higher optical depth and subsequently higher temperature fluctuations. 

\section{Secondary CMB fluctuations from a distribution of quasars.}
\label{sec:kSZ_quasar}

\begin{figure}
\centering 
\includegraphics[width=\columnwidth]{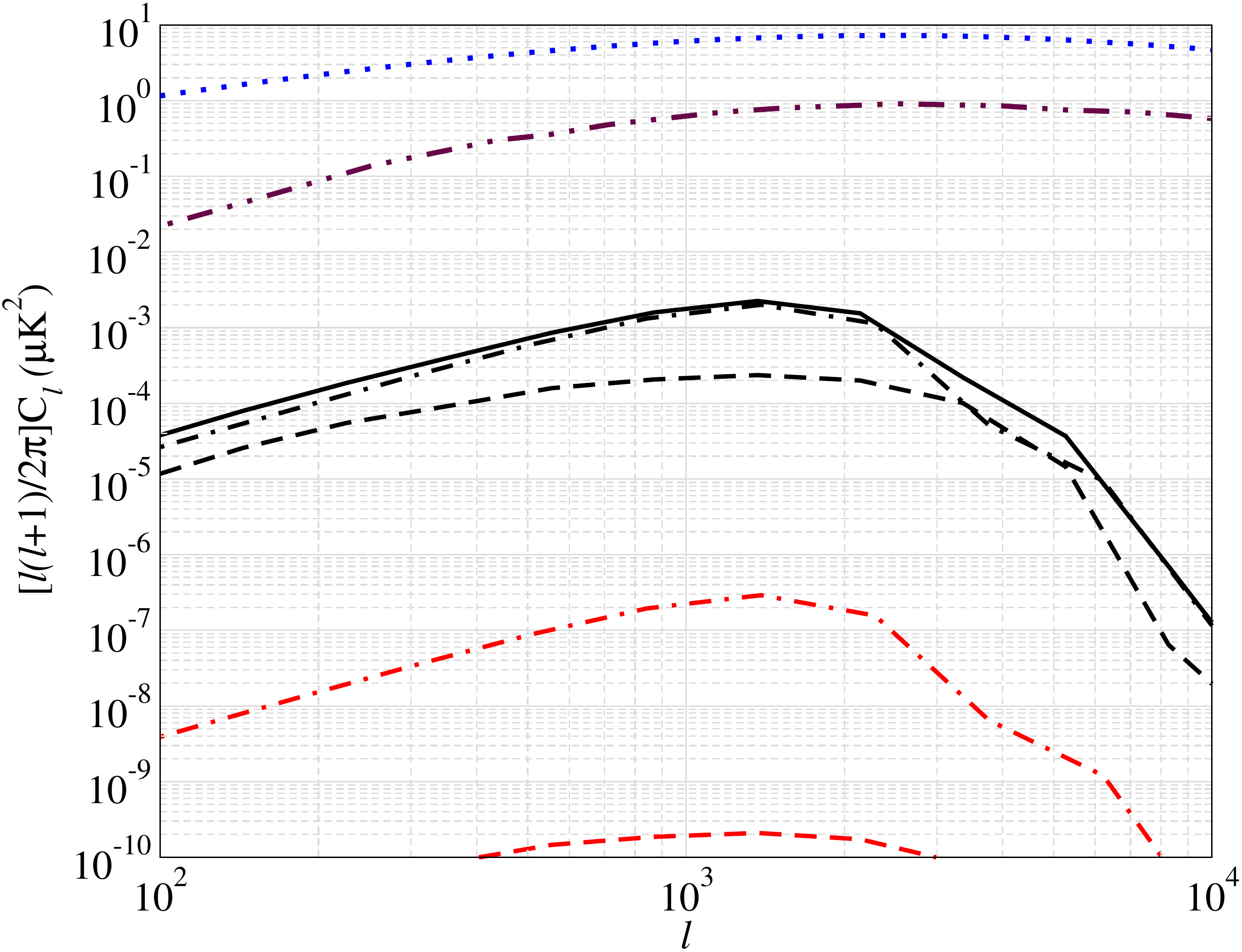}
\caption{Power spectrum from kSZ due to ionized bubbles around quasar with halo model calculations ($f_Q=0.01$) (red), and using the QLF from \citep{SHFARRH2020} (black).One-halo, two-halo terms and the total are plotted in dot-dashed, dashed and solid lines respectively. Since the bubbles have sharp edges which leads to oscillations after Fourier transform, we plot binned power spectrum by combining logarithmically spaced $\ell$ bins. The quasar lifetime is $10^7$ yr and $\dot{N}=10^{56}$ s$^{-1}$. We also show tSZ from galaxy clusters \citep{BCKM2018} but using our own calculation as a reference (dotted blue line) and reionization kSZ \citep{CMB_HD} in maroon double dot-dashed line. Note that in this figure, the bubble sizes were computed assuming the universe to be neutral.
}
\label{fig:quasar_Cl}
\end{figure}

Given the \ksz distortion from any single quasar bubble and the space density of high redshift quasars in hand, it is straight forward to calculate the secondary CMB fluctuations due to these bubbles using halo model \citep{S2000}.
 The expression for kSZ  from a single bubble in multipole space, after fourier transforming from angular space ($y(\theta)=\frac{\Delta T(\theta)}{T_{\rm CMB}})$, is given by \citep{KS2002,HP2013},
 \begin{equation}
 y_\ell(M,z)=\sigma_{T} n_e(z)\frac{v_r}{c}\frac{4\pi R(z)}{l_R ^2}\int dx x^2\frac{sin(\ell x/l_{R})}{\ell x/l_{R}}\rho(x),
 \end{equation}
where  $R(z)$ is the size of bubble, $D_A(z)$ is the angular diameter distance, $l_R=D_A(z)/R(z)$, $x=r/R(z)$, $n_e(z)$ is the background electron density, $v_r(z)$ is the rms radial cosmic velocity at the bubble location, and $\rho(x)$ is the normalized density distribution of electrons inside a bubble.
 
  In the linear regime, applicable at physical scales corresponding to the quasar bubbles, the rms velocity has a simple form, $v_r(z)=v_r(0)(1+z)^{-1/2}$ \citep{Peebles1980,ADPG1996}  with $v_r(0)\approx 0.001c$ which can be computed from linear perturbation theory, and using the publicly available code {\tt CLASS} \citep{BLT2011}. For kSZ, only the radial term contributes, thus $v_r(0)=\frac{0.001c}{\sqrt{3}}$  Note that there are detection of bulk velocities at low redshifts \citep{ACT2021ksz} but not for high redshifts. Measuring high-$z$ bulk velocities remain one of science goals of future CMB experiments as well as the Rubin Observatory \citep{LSST2019}. In absence of such high-$z$ measurements, we have used theoretically well motivated linear approximation for our high-$z$ bulk flows. As a first approximation, we can assume all the quasars to have the  same luminosity such that $\dot{N}=10^{56}$s$^{-1}$ and also the same radius $R(z)=R_0(1+z)^{-1}$ with $R_0\approx 10$ Mpc. This is assuming all the hydrogen to be neutral which is, of course, not the case during the reionization. We will return to this point later. 
  
In the analytical halo model approach, the kSZ power spectrum can be written as the sum of the one-halo and the two-halo terms. The expressions for one and two-halo terms analogous to  galaxy clusters are given by \citep{MB2000,M2001,KS2002},
\begin{equation}
C_\ell^{1-{\rm halo}}=g_{\nu}^2\int_{10}^{z_{\rm lim}} {\rm d}z \frac{{\rm d}^2V}{{\rm d}z{\rm d}\Omega}\int {\rm d}M f_Q\frac{{\rm d}n(M,z)}{{\rm d}M}|y_\ell(M,z)|^2,
\label{eq:onehalo_kSZ}
 \end{equation}
 where  $g_{\nu}$ is a frequency-dependent parameter which is equal to 1 for kSZ due to its blackbody spectrum, $\frac{d^2V}{dzd\Omega}$ is the comoving volume element per steradian and $\frac{{\rm d}n(M,z)}{{\rm d}M}$ is the comoving number density of dark matter halo. Similarly, the two-halo term can be written as,
 \begin{equation}
   \begin{split}
   C_\ell^{2-{\rm halo}}=g_{\nu}^2\int_{10}^{z_{\rm lim}} {\rm d}z \frac{{\rm d}^2V}{{\rm d}z{\rm d}\Omega}\left[\int {\rm d}M f_Q\frac{{\rm d}n(M,z)}{{\rm d}M}b(M,z)y_\ell(M,z)\right]^2 \\ P_{\rm lin}\left(\frac{\ell}{\chi(z)}\right),
\label{eq:twohalo_kSZ} 
\end{split}
 \end{equation}
 where $P_{\rm lin}(k)$ is the cosmological matter power spectrum in the linear regime and 
 $b(M,z)$ is the dark matter halos bias \citep{TRKKWYG2010} given by
 $b(M,z)=1+ \left(\frac{\delta^2_c}{\sigma(M,z)^2}-1 \right) / \delta_c$
with $\delta_c=1.686$.  

The redshift integral is between $z_{\rm lim}=6$ to 10. The integral is not sensitive to the higher bound of redshift as the halo mass function exponentially dies off at higher redshifts while it is very sensitive to the lower end which is the reionization redshift at which point all the bubbles merge together to form a uniform pattern. 
  We have taken integral over halo mass $M_{min}$ to infinity with $M_{min}=10^{12}M_{\odot}$.  In all our power spectrum plots, we have averaged over $\ell$-bins at high $\ell$ such that there are no ringing patterns due to a sharp cutoff in $\rho(x)$ at the bubble boundary as seen in Fig. \ref{fig:xHI_evolution}.
 
To compute the kSZ power spectrum using the QLF of \cite{SHFARRH2020}, we redo a halo model calculation but adding an artificial boost factor (analogous to an halo occupation distribution with) term in front of halo mass function i.e, we choose a factor $B(z)$ such that,
\begin{equation}
B(z)\times\int {\rm d}M \frac{{\rm d}n}{{\rm d}M}=n_Q(z),
\label{eq:hod_proxy}
\end{equation}
where $n_Q(z)$ is the number density of quasars as in Fig. \ref{fig:aghanim_vs_halo_model}.\\

Putting everything together, in Fig. \ref{fig:quasar_Cl}, we plot the 1-halo and 2-halo \ksz power spectrum from our halo model calculations and compare them with more accurate QLF results of \cite{SHFARRH2020}. It is clear that halo model with $f_Q=0.01$ grossly under-estimate the $C_\ell$s compared to accurate QLF which is directly related to their  predicted abundance of quasars  as shown in Fig. \ref{fig:aghanim_vs_halo_model}.  At high redshifts, due to exponential suppression of halo mass function, the mass integral is dominated by halos at the limiting mass $M_{min}$. This suppression also overwhelms the increase coming from the heavier halos being more biased. Therefore, 2-halo term is close to the 1-halo term but modulated by linear matter power spectrum and with a multiplicative factor $b(M_{\rm min})^2$. 

In \cite{ADPG1996}, the fiducial value of $\dot{N}$ was chosen to be $10^{57}$s$^{-1}$. This makes the sizes of bubbles bigger. As can be seen from the expression of $y_{\ell}$, there is a $1/\ell^2$ dependence. This is due to the phase cancellation of fourier modes following a line-of-sight integration\footnote{For a discussion, see \cite{HW1997}}. For $\dot{N} = 10^{57}$s$^{-1}$, this results in an increase in the contribution from quasars to the total kSZ $C_{\ell}$, which for low $\ell$-values can become comparable or even greater than total reionization kSZ $C_{\ell}$ found in simulations (for example, by \citet{SBDHHLOT2010}). For more more realistic $\dot{N}$,  quasar contribution being smaller, our calculations confirm the galaxies to be the main driver of reionization at all multipoles. 


\subsection{Alive vs dead quasar}

\begin{figure}
\centering 
\includegraphics[width=\columnwidth]{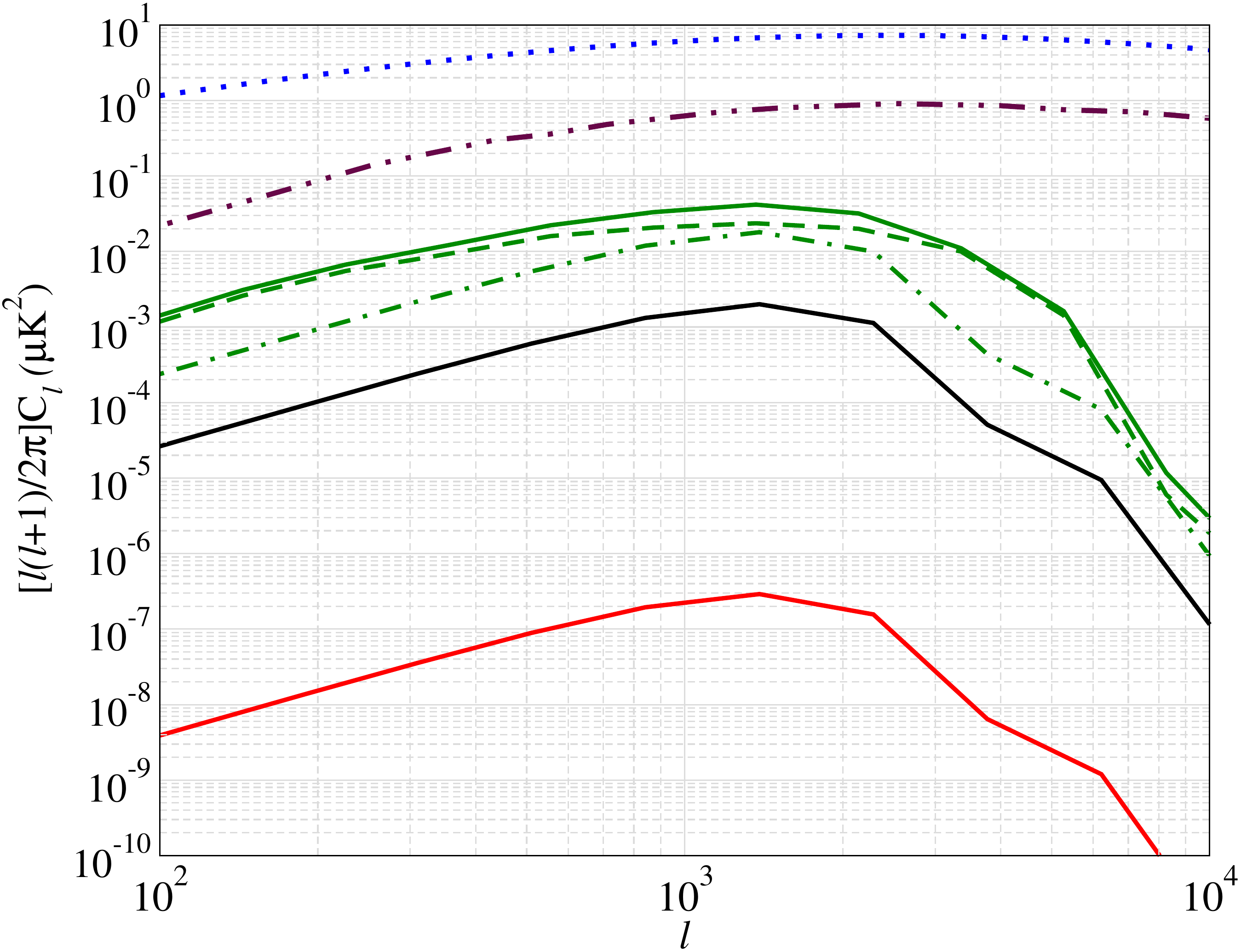}
\caption{Same as in Fig. \ref{fig:quasar_Cl}, but now including dead quasars shown with green lines.
}  
\label{fig:quasar_Cl3}
\end{figure}

While Eq. \ref{eq:number_density} gives the number density of active quasars, it has been shown that the fossil ionized bubble around dead quasars can survive long after the quasar engine shuts off \citep{FHO2008}. We take this into account by re-defining the  number density of all quasars as,
\begin{equation}
n^{\rm eff}_Q(z)=\int dt \frac{n_Q(z)}{t_Q},
\end{equation}
where $t$ is the cosmic time at redshift z and $n_Q(z)/t_Q$ is the rate of production of quasars. We then use Eq. \ref{eq:hod_proxy}, replacing $n_Q(z)$ with $n_Q^{\rm eff}(z)$.

In Fig. \ref{fig:quasar_Cl3}, we replot the kSZ power spectrum including contributions from dead quasars. The first point to notice is that, in the presence of dead quasars, the 2-halo term can become larger than the 1-halo term. In comparison, with live quasars only, the two halo term is either similar or smaller to the one halo contribution (see Fig. \ref{fig:quasar_Cl}). The underlying reason for this difference comes from the two halo term being proportional to $f_Q^2$ while the one halo term is proportional to $f_Q$.  Thus, if $f_Q$ is relatively large, the two halo term dominates.

 It is also seen that the resultant  kSZ $C_{\ell}$ can be a percent of the  reionization kSZ power spectrum at $l\sim 10^3$, but can become significant ($\sim$ 10\%) at  $\ell$ $\sim$ 100. 
Since future CMB experiments aimed at probing reionization are geared towards high $\ell$s, we do not expect kSZ from quasars bubbles to introduce any significant bias in patchy reionization measurements. Also, we have made an important assumption in these calculations which is that the universe is neutral upto at least $z\approx 6$. Clearly, this is a very simplistic assumption and is not valid within our current understanding of the Universe \citep{Planckreion}. In the next section, we take into account the reionization history of universe while computing the kSZ power spectrum.


\begin{figure}
\centering 
\includegraphics[width=\columnwidth]{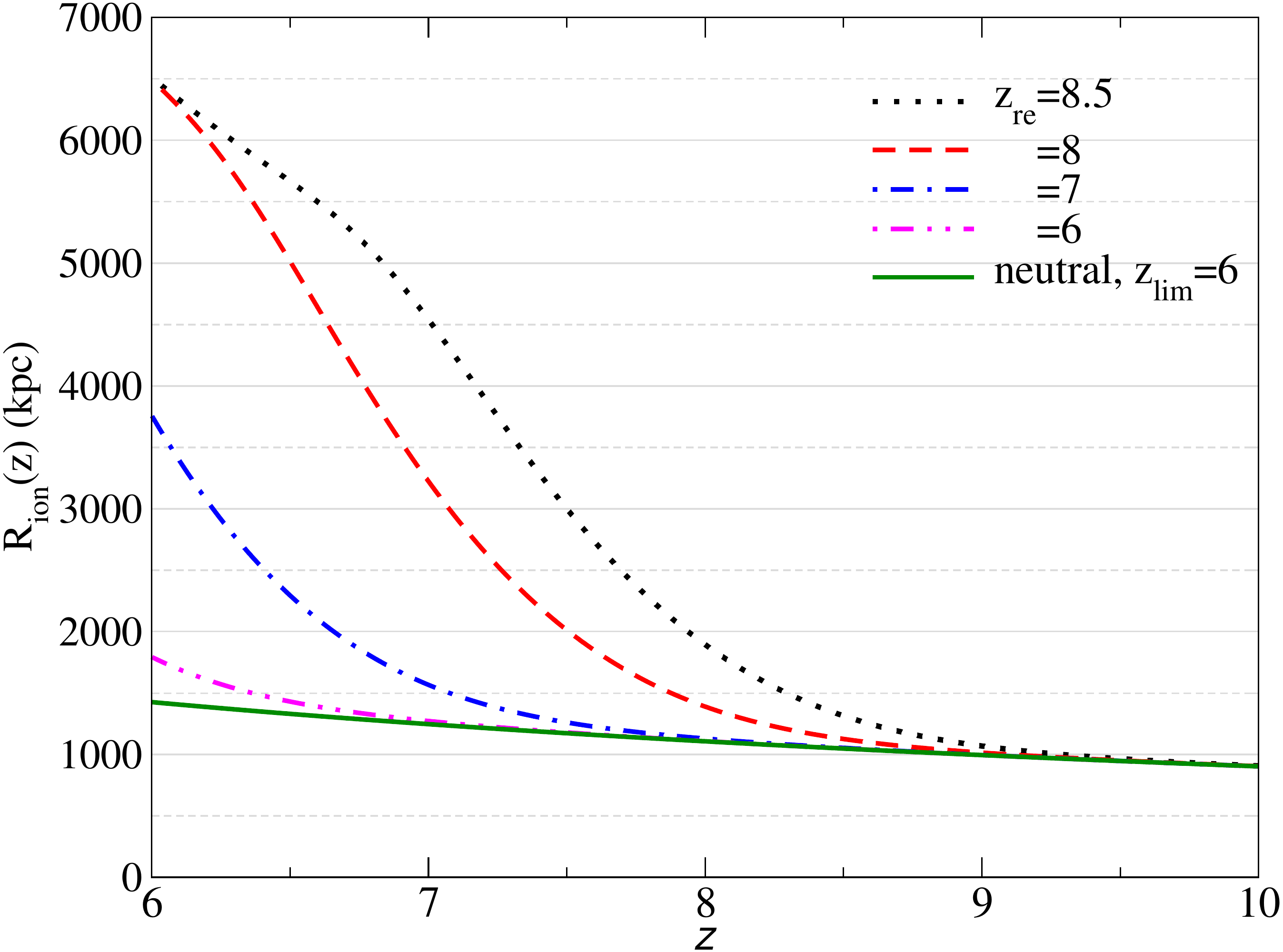}
\caption{Evolution of the bubble size in a neutral universe (black) and a universe with evolving mean electron fraction  given in Eq. \ref{eq:reionization}. The label $z_{re}$ shows  the mid point of reionization. For our kSZ power spectrum calculations, we integrate only upto $z_{re}$ and we do not use the information below this redshift where one needs to take into account merging of bubbles in an highly ionized universe. 
}  
\label{fig:bubblesize_vs_z}
\end{figure}

\begin{figure}
\centering 
\includegraphics[width=\columnwidth]{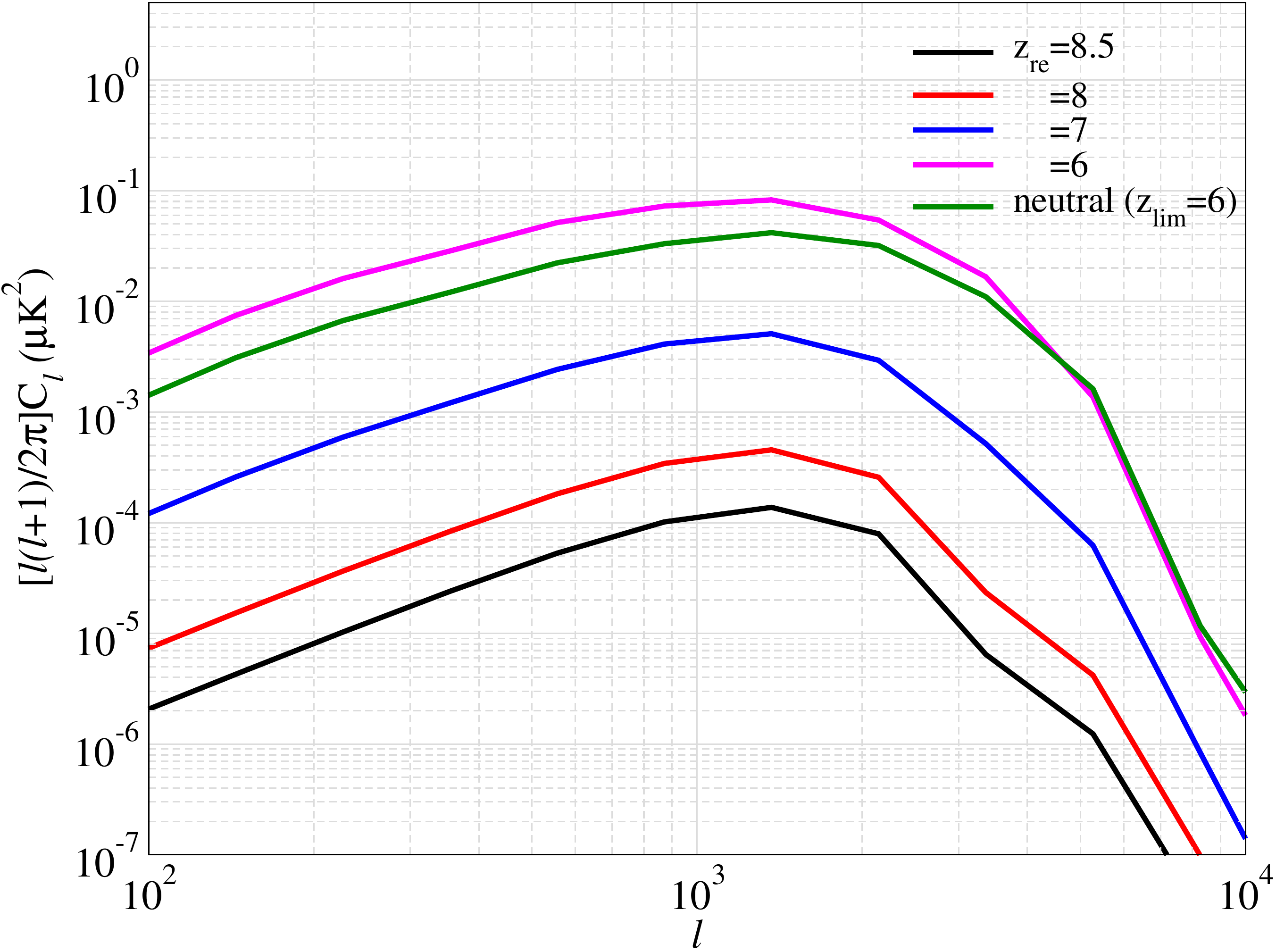}
\caption{ The kSZ power spectrum from CMB distortion by ionized bubbles with different reionization history, $z_{re}$ = \{8.5, 8, 7, 6\} (see text for details) . The case where our Universe remains neutral till $z=6$ is also shown.
}
\label{fig:kSZ_Cl_reion}
\end{figure}

\begin{figure}
\centering 
\includegraphics[width=\columnwidth]{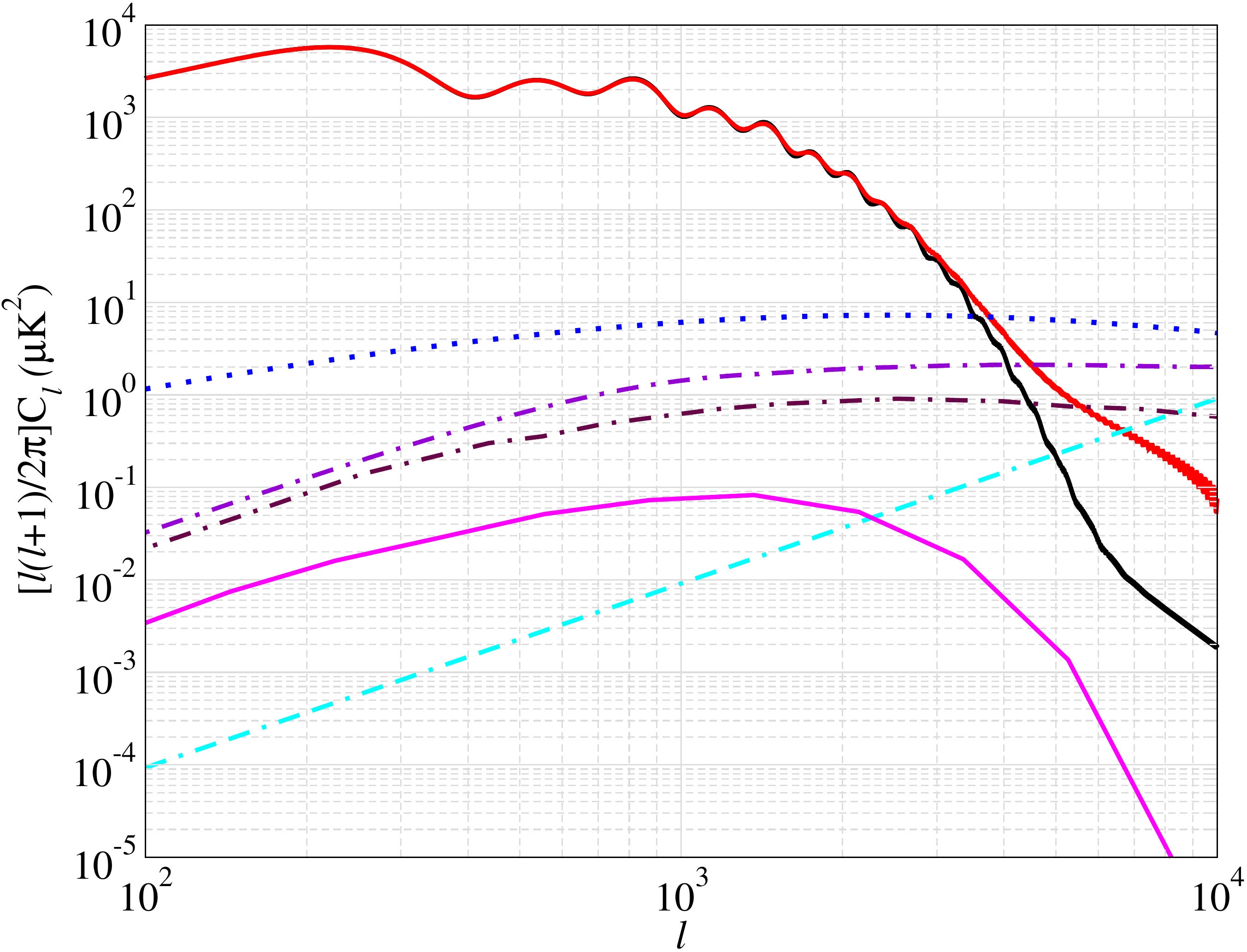}
\caption{The CMB fluctuation power spectrum from primary unlensed CMB (solid black), lensed cmb (solid red), thermal SZ from clusters (dotted blue), total kSZ including late kSZ from clusters (dot-dashed violet), reionization kSZ (dot-dashed maroon) and quasar bubbles for reionized universe with $z_{re}\approx 6$ (solid magenta) are shown. The template for kSZ is derived from simulations \citep{SBDHHLOT2010} . The instrumental noise expected for the future CMB-HD survey ($0.5 \mu$K-arcmin) is also shown (dot-double dashed cyan straight line). 
 }  
\label{fig:Cl_template}
\end{figure}

\section{Dependency of kSZ $C_{\ell}$ on reionization}

Irrespective of our incomplete understanding of reionization, we can use representative reionization models to compute the quasar bubbles for more realistic scenarios. In this work, we consider redshift-symmetric reionization model which satisfy the optical depth constraint on reionization from CMB observation \citep{Planckreion}. In this model, the onset of reionization (assuming the universe to be composed of only hydrogen) is given the redshift dependent ionization fraction
\begin{equation}
    x_e(z)=0.5\left[1+ {\rm tanh}(y-y_{re})/\delta y)\right],
    \label{eq:reionization}
\end{equation}
where $y=(1+z)^{3/2}$ and $\delta y=\frac{3}{2}(1+z)^{1/2}\delta z$.
 This model gives us a period of extended reionization with the reionization redshift, $z_{re}$,  defined as the redshift at which mean ionization fraction in the universe becomes 50\%. There are further constraints on the duration of reionization using kSZ observations from SPT-SZ + SPTpol Surveys \citep{SPTksz2021}  which looks at the redshift interval within which the mean ionization fraction changes from 10\% to 90\%. However, these constraints currently depend upon dividing the total kSZ between a patchy and a homogenous reionization components. There are further constraints from Lyman-alpha observations \citep{KKHBPCA2019} which indicate that reionization was completed as late as $z\lesssim 5.5$. In summary, our choice of the reionization model is not unique in anyway and the kSZ power spectrum will be very much dependent upon the reionization history. However, a later reionization boosts the kSZ power and hence would be favourable for a detection of the \ksz from quasar bubbles.  In this work, we look at $z_{re}\,=\,\{8.5,8,7,6\}$ and  $\delta z=0.7$.

Another source of uncertainty in our calculations is the amount of CMB distortions from these bubbles as we reach the redshift chosen as the end point of reionization. At the start of reionization, the bubbles are far apart giving rise to the temperature fluctuations computed here. As reionization proceeds, these bubbles start to merge and fluctuations start to disappear. However, it is unphysical to think of a single redshift where the Universe gets reionized fully. As a conservative approximation, we can calculate the secondary anisotropies generated only upto $z_{re}$  and use this redshift as the choice for the integration limit in Eq. \ref{eq:onehalo_kSZ} and \ref{eq:twohalo_kSZ}.  Any additional CMB anisotropy generated post $z_{re}$ will only add to the numbers quoted in this work.

There is one more observable importance of a late reionized Universe.
As shown in Fig. \ref{fig:bubblesize_vs_z}, a quasar bubble grows (Eq. \ref{eq:Ifront_analytic}) to a larger size in an ionized medium, compared to a neutral medium. Hence,  it is relatively easier to resolve and observe quasar bubbles near the end of reionization. However, one has to keep in mind that it can be difficult to find an individual intact large bubble at lower redshifts due to rapid overlapping of ionized bubbles. Note, that as reionization proceeds, the overlap of the bigger bubbles significantly reduces the temperature aniostropy. High resolution numerical simulations are needed to probes CMB anisotropies near the end of the reionization.

We compute the kSZ power spectrum (shown in Fig. \ref{fig:kSZ_Cl_reion}) using growth of the quasar ionized bubbles and their distribution as described till now. We begin with the lower redshift integral $z_{lim}$ fixed at $z_{re}=8.5$, and then keep lowering $z_{lim}$ upto $z_{re}=6$. The reionization redshift has a very strong impact on the amplitude of the \ksz power spectrum which can vary by orders of magnitude. This is expected due to steep dependence of the QLF as a function of redshift (Fig. \ref{fig:aghanim_vs_halo_model}). 

The strength of the kSZ power spectrum (as in  Fig. \ref{fig:kSZ_Cl_reion}) can be shown to scale with the reionization redshift as
\begin{eqnarray}
{\cal D}^{\rm 1-halo}_{\ell = 1500} &=& 1.35\times 10^{-4}(1.0+z_{re})^{-1.09z_{re}+9.27}\\ \nonumber
{\cal D}^{\rm 2-halo}_{\ell = 1500} &=& 1.73\times 10^{-6}(1.0+z_{re})^{-2.13z_{re}+18.1} \; ,
\end{eqnarray}
where ${\cal D}_\ell = \frac{\ell (\ell +1)}{2\pi} C_\ell$. The change in the location of the peak of the power spectrum, with varying $z_{\rm re}$, is negligible since (i) the comoving distance changes very slowly with redshift and hence $\ell_R$ does not change much, and (ii) binning in $\ell$ further washes out fine structural details. Note, however, that between a neutral ($z_{\rm lim} = 6$) and a $z_{\rm re}=6$ universe, bubbles grow larger in the ionized universe and the resultant power spectrum peaks a little earlier in $\ell$-space.

The peak of the kSZ $C_\ell$ also moves to lower $\ell$ as the bubble sizes increase as seen in Fig. \ref{fig:bubblesize_vs_z} which subtends bigger angle on the sky. 
The quasar $C_{\ell}$s are 1-2 orders of magnitude smaller than the tSZ from galaxy cluster and late time kSZ as shown in Fig. \ref{fig:Cl_template}. However, we remind the reader that in the Fig. we have chosen the optimistic case with reionized universe with $z_{re}\approx 6$. In reality, the quasar kSZ $C_{\ell}$s are expected to be order of magnitude smaller as shown in Fig. \ref{fig:kSZ_Cl_reion}. To compute kSZ from reionization, one needs to do simulations \citep{MFHZZ2005,IPBMS2007,MMS2012,SRN2012,BTCL2013,A2016}. In these simulations, one typically assumes that galaxies within dark matter halos having mass $10^{9}-10^{10}$ M$_\odot$ give rise to reionization-kSZ and the magnitude of kSZ fluctuations turns out to be of the same order as late kSZ (FIg.1 of \citep{CMHK2021}. Our calculations show that, quasars contribute a negligible amount to reionization-kSZ with the accurate QLF of \cite{SHFARRH2020}. Note, that the quasar bubble $C_\ell$s have a peak at $\ell \gtrsim 2000$ whereas the $C_\ell$ due to patchy reionization flattens off at $\ell > 1000$. Hence, by measuring $C_{\ell}$ over multiple $\ell$-bins, we need accurate $C_{\ell}$ templates to distinguish one source from other \footnote{This technique is already used in \cite{SPTksz2021} for the detection of reionization-kSZ.} with a futuristic survey like CMB-HD. Apart from measuring the power spectrum. 
we may also have good prospects at detecting individual large ionized bubble as shown in the next section.

\begin{figure}
\centering 
\includegraphics[width=\columnwidth]{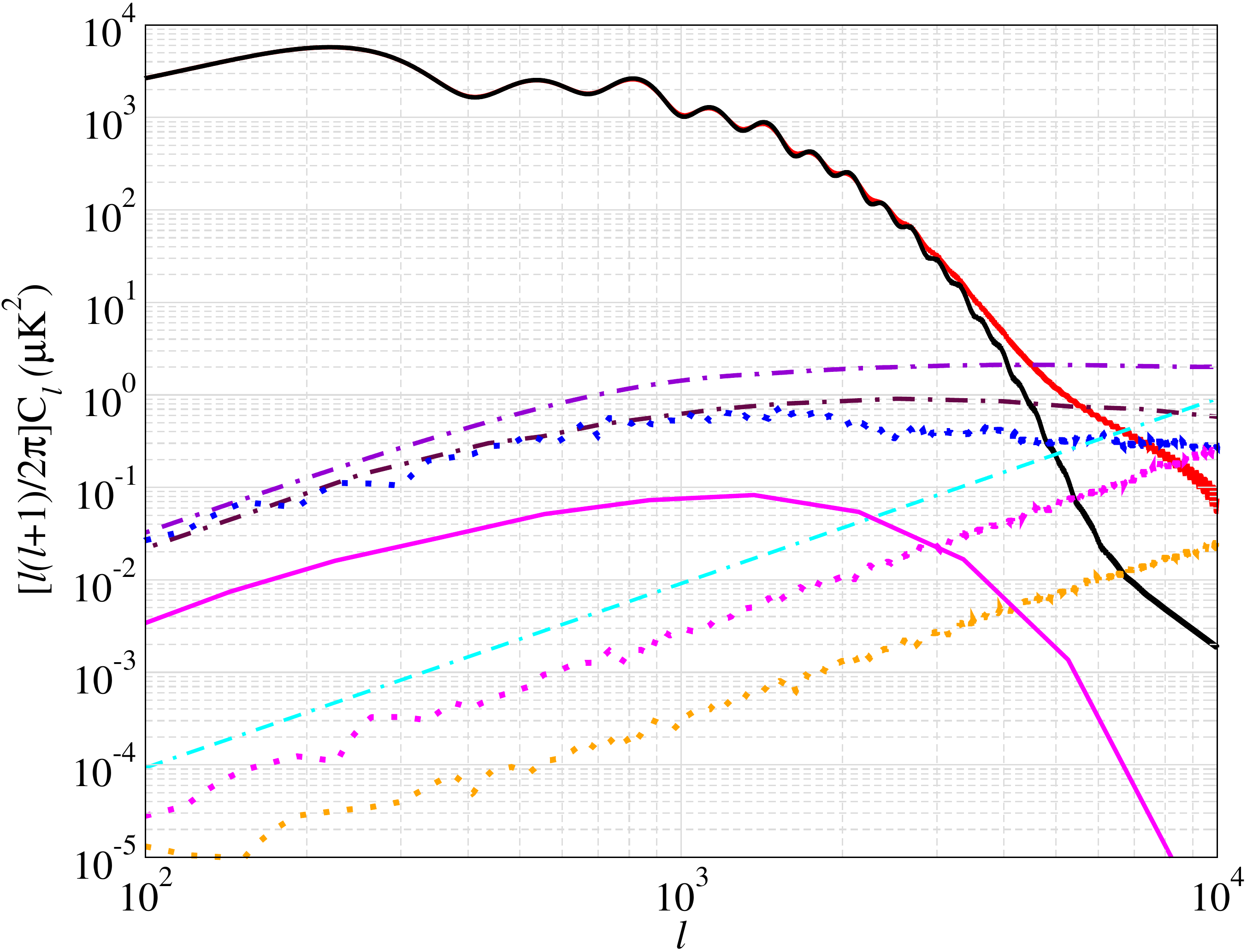}
\caption{Same as in Fig. \ref{fig:Cl_template} but for a foreground-reduced map (see subsection \ref{subsec:foreground} for details). Note the new foreground-reduced  tSZ (dotted blue) and the additional foregrounds from IR point sources (dotted magenta) and radio point sources (dotted orange). 
 }  
\label{fig:fg_removal}
\end{figure}

\begin{figure}
\centering 
\includegraphics[width=\columnwidth]{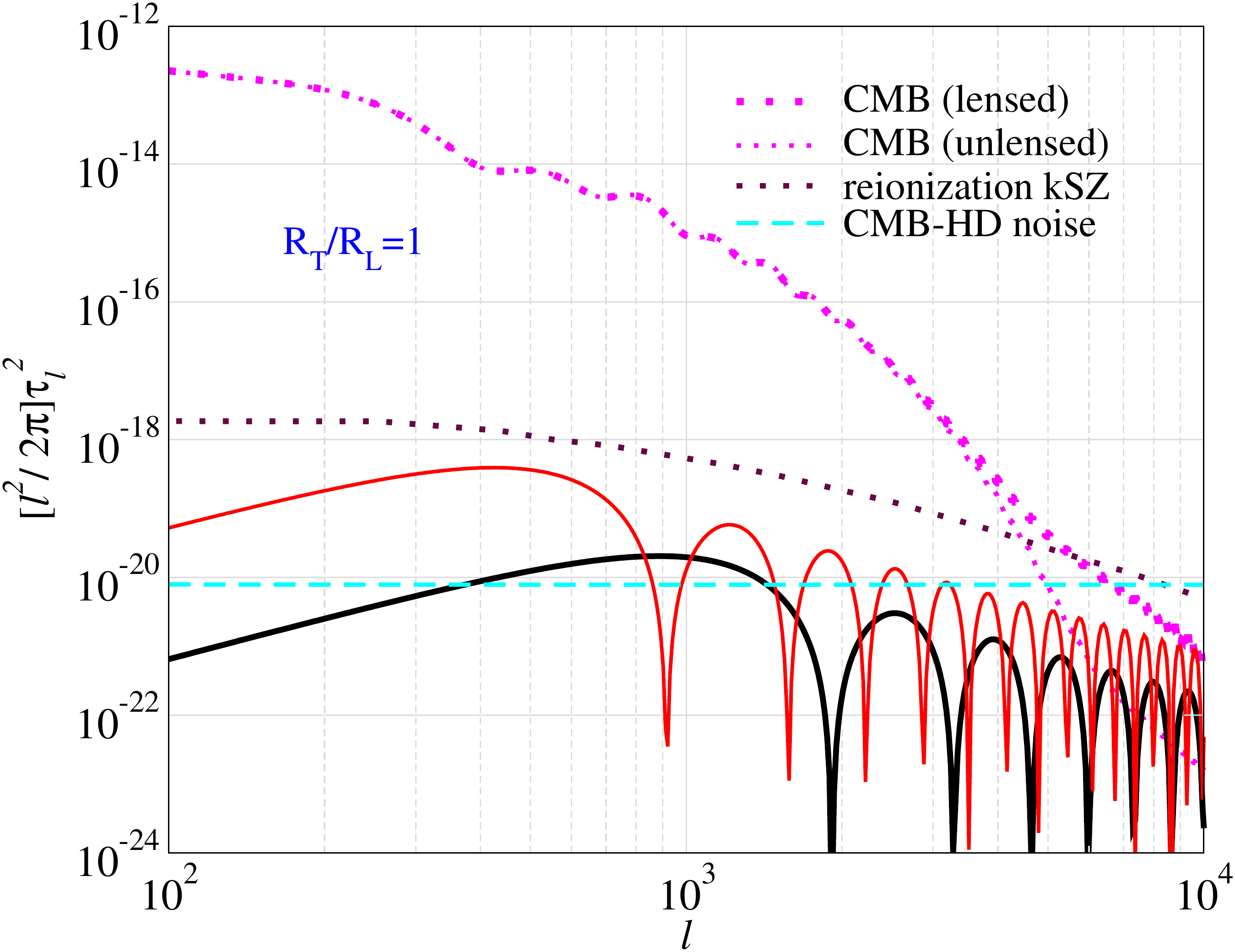}
\caption{The single quasar bubble anisotropy power spectrum $\frac{\ell^2}{2\pi}\,\tau_{\ell}^2$, where $\tau_{\ell}$ is the Fourier transformed spatial temperature anisotropy profile from a single quasar bubble located at $z = 8$. We have assumed $\dot{N}=10^{57}$s$^{-1}$ with $f_H=1$ and two different quasar lifetimes of $t_Q=10^7$ (thick black line) and $10^8$ yrs  (thin red line) respectively. The $C_{\ell}$ from individual bubbles depends on the size of the bubble, and therefore is completely degenerate with $\dot{N}$ and $t_Q$. We also show lensed and unlensed primary CMB anisotropy power spectrum ($C_{\ell}$s), reionization kSZ and the noise $C_{\ell}$s from a CMB-HD-like \citep{CMB_HD} experiment. 
}
\label{fig:Cl_comp}
\end{figure}

\section{Direct Detection of Quasar Bubbles}
\label{sec:matched_filter}
In this section, we take the first step at the prospects of extracting the kSZ signal of an individual quasar bubble from future  lensed CMB maps, thus detecting a single bubble at a particular redshift. Unfortunately, multifrequency subtraction such as Internal linear combination \citep{TTH1997,HP2013} cannot be used to distinguish between the kSZ signal from the bubble and the primary CMB since they have the same spectral signature.
 Therefore, we need to distinguish the spatial signature of a quasar bubble to separate  it from the primary CMB fluctuations. While the CMB is dominant at large scales ($l\lesssim 1000$), the fluctuations damp out exponentially at smaller scales; in contrast, the Fourier signature of the kSZ profile from a bubble can extend to high $\ell$'s.  Under the assumption that we can model the kSZ profile of the individual quasar bubble, one can  apply the matched filtering approaches to extract the bubble signature from the CMB. However, the procedure is even more complicated due to the presence of tSZ from galaxy clusters, and foregrounds such as infrared and radio point sources. At high $\ell$s these foregrounds can swamp even the dominant primary CMB  (Fig. 11 of \cite{SBDHHLOT2010}) and, hence, need be removed to a sufficient level before we can search for kSZ from quasars.
 
\subsection{Foreground-reduced CMB maps}
\label{subsec:foreground}
 
We \citep{Mankar} closely follow the procedure of \cite{HD2022} for the foreground removal. We extract a patch of sky from the simulation \citep{SBDHHLOT2010} at a frequency of 150 GHz. Assuming that the IR and radio point sources can be detected at $5\sigma$ level for a CMB-HD  type of multi-frequency experiment \citep{CMB_HD}, one can use a masking template to remove the point sources.   This amounts to a fluxcut of $\sim 0.03$mJy for both radio and IR sources at 150 GHz. For the IR point sources, we take into account uncertainty in source subtraction using the same procedure as in \cite{HD2022}. We also mask out galaxy clusters which will be detected at 5$\sigma$ level in a \cite{CMB_HD}-like experiment with redshift dependent mass threshold (Fig. 3 of \cite{Raghunathan2022}) and using the halo catalog of \cite{SBDHHLOT2010}. We cut out a circle of size of 10 arcminutes if the galaxy cluster is located at $z\lesssim 0.5$ or 5 arcminutes if vice versa. It can be seen from Fig. \ref{fig:fg_removal},  that the foregrounds are  small compared to CMB and patchy reionization kSZ, and the main idea is to have the foregrounds reduced further to a level such that only CMB, kSZ and the instrumental noise dominate. However, the quantities obtained in the figure are sky-averaged quantities and the kSZ power spectrum from a distribution of quasar bubbles is still dominated by the residual foregrounds. For an individual bubble detection, one has to work on a map level and apply match-filtering to a map containing the quasar bubbles to study its detection prospects. This aspect is currently under investigation \citep{Mankar}. Here, we assume an optimistic scenario where CMB, kSZ and instrumental noise ($0.5 \mu$K-arcmin) are the main source of noise for bubble detection. In a futuristic survey, one may also subtract late kSZ component as these are caused by galaxy clusters which produce tSZ signal too. Therefore, we only consider contribution from patchy reionization kSZ for a first  optimistic estimate of detection prospects of kSZ from quasar bubbles. 

\subsection{Detection prospects}

The kSZ spatial template of a single quasar bubble is dictated by the free electron density profile around the quasar (Sec. \ref{subsec:radiative_transfer}, Fig. \ref{fig:xHI_evolution}) given that the optical depth is proportional to electron number density along  the line-of-sight through the ionized medium.  Even in the presence of any gradient/sub-structures  in free electron density, the medium is practically completely ionized till the bubble boundary. Therefore, to a first approximation, the density profile can be taken to be a box function with sharp edges. While electron number density is constant within the bubble, the observed angular dependence of the path through the bubble is proportional to $\sqrt{\theta_0^2-\theta^2}$, where $\theta_0$ is the angular width of the object subtended on the sky for an observer.

In the previous sections, we chose our fiducial quasar parameters to be $\dot{N}=10^{56}$s$^{-1}$, $f_H=1$ and $t_Q=10^7$ yr. The temperature fluctuations from these bubbles are of the order $\sim 10^{-7}$ (Fig. \ref{fig:quasar_optical_depth}). We see from Fig. \ref{fig:fg_removal} that the lensed CMB+kSZ+instrumental noise would be of the order of $\sim$  few $\mu K$. Therefore, bigger bubbles with higher temperature fluctuations are more likely to be found in a CMB map. As a first attempt, for exploratory purpose, we choose $\dot{N}=10^{57}-10^{58}$s$^{-1}$ and $t_Q=10^7-10^8$ yr. We also choose bubbles with a surrounding neutral medium so that there is a well-defined bubble boundary. To satisfy this criteria, we assume the bubble to be around $z\approx 8$, the redshift at which around 50 percent of the universe is expected to be reionized \citep{Planckreion}. However, we vary $f_H$ in order to understand the sensitivity of our calculations to the details of the reionization. The angular size of bubbles at this redshift and parameters would correspond to $\ell\sim 1000$. At such small angular scales, we can ignore the curvature of the sky and simply Fourier transform between $\ell$ and $\theta$.

The signal to noise ratio, in the case of matched filtering, is given by \citep{HT1996,MBD2006}, 
\begin{equation}
S/N=\sqrt{\frac{1}{(2\pi)^2}\int d^2k \frac{\tau(k)^2}{P_k}},
\label{eq:matched_filter}
\end{equation} 
where $\tau(k)$ is the Fourier transformed temperature anisotropy profile of the source (here, the quasar bubble) convolved with the beam and $P_{k}$ is the total noise. We assume that the only sources of noise are the primary CMB fluctuations, patchy reionization kSZ and the instrumental noise. In the flat sky approximation, there is one-to-one correspondence between Fourier mode $k$ and spherical harmonic mode $\ell$. Therefore, we replace 
$\tau(k)$ with $\tau_{\ell}$ and $P(k)$ with $C_\ell$ of CMB, reionization kSZ and instrumental noise in Eq. \ref{eq:matched_filter}.   We obtain the power spectrum ($C_{\ell}$) of the primary CMB using the publicly available code {\tt CLASS} \citep{BLT2011} and compare it to the kSZ power from both reionization and from a single fiducial quasar bubble in Fig. \ref{fig:Cl_comp}.
 One can see that the kSZ signal starts to gain on CMB at higher $\ell$s. It is clear that we need arcminute resolution experiments capable of probing $l\approx 10^4$  to distinguish the two signals. As a future possibilty, we consider proposed CMB-HD \citep{CMB_HD} experiment which expects to have arcminute resolution with instrumental noise as shown in Fig.\ref{fig:Cl_comp}. In the presence of CMB $C_{\ell}$ (lensed), reionization kSZ and CMB-HD noise, we obtain a signal-to-noise ratio of 0.08 and 0.17 for the two cases shown in Fig. \ref{fig:Cl_comp}. In estimating the S/N, we
 have assumed the beam to be sufficiently sharp to be approximated as a delta function which is reasonable assumption in our case. We divide the multipole range $10^2<\ell<10^4$ in 100 linearly spaced bins. Extending to $l>10^4$ does not help since the kSZ power goes down at smaller scales  while the instrumental noise starts to exponentially rise once we are at scales below instrumental resolution.
 
\begin{figure}
\centering 
\includegraphics[width=\columnwidth]{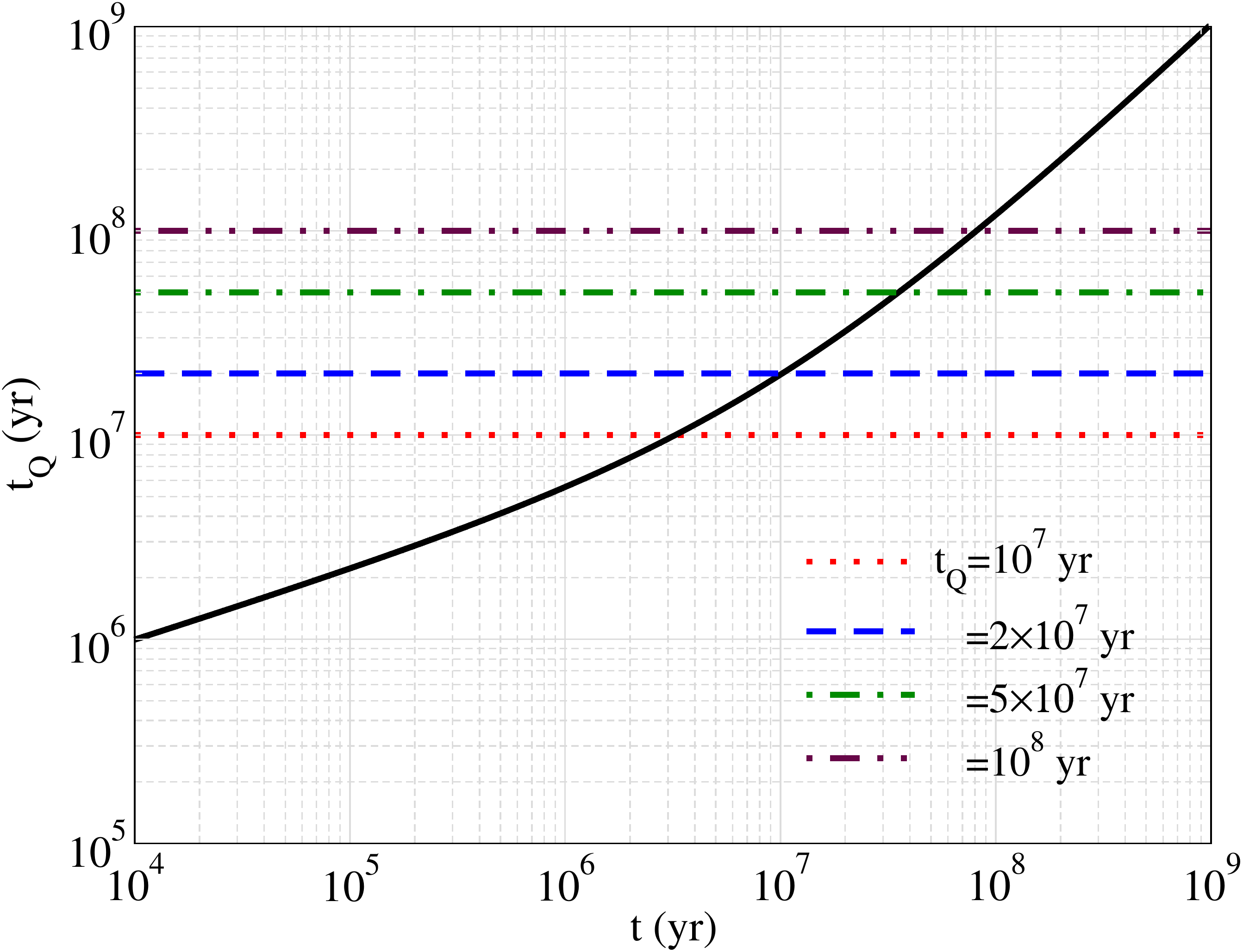}
\caption{Comparison of apparent age of the quasar with its true age in transverse direction relative to line of sight. The black line corresponds to $\tau+\frac{R(\tau)}{c}$. Luminosity of quasar is $10^{57}$s$^{-1}$ at $z=8$ with $f_H$ assumed to be 1.  
}
\label{fig:distortion_shape}
\end{figure}

\begin{figure}
\centering 
\includegraphics[width=\columnwidth]{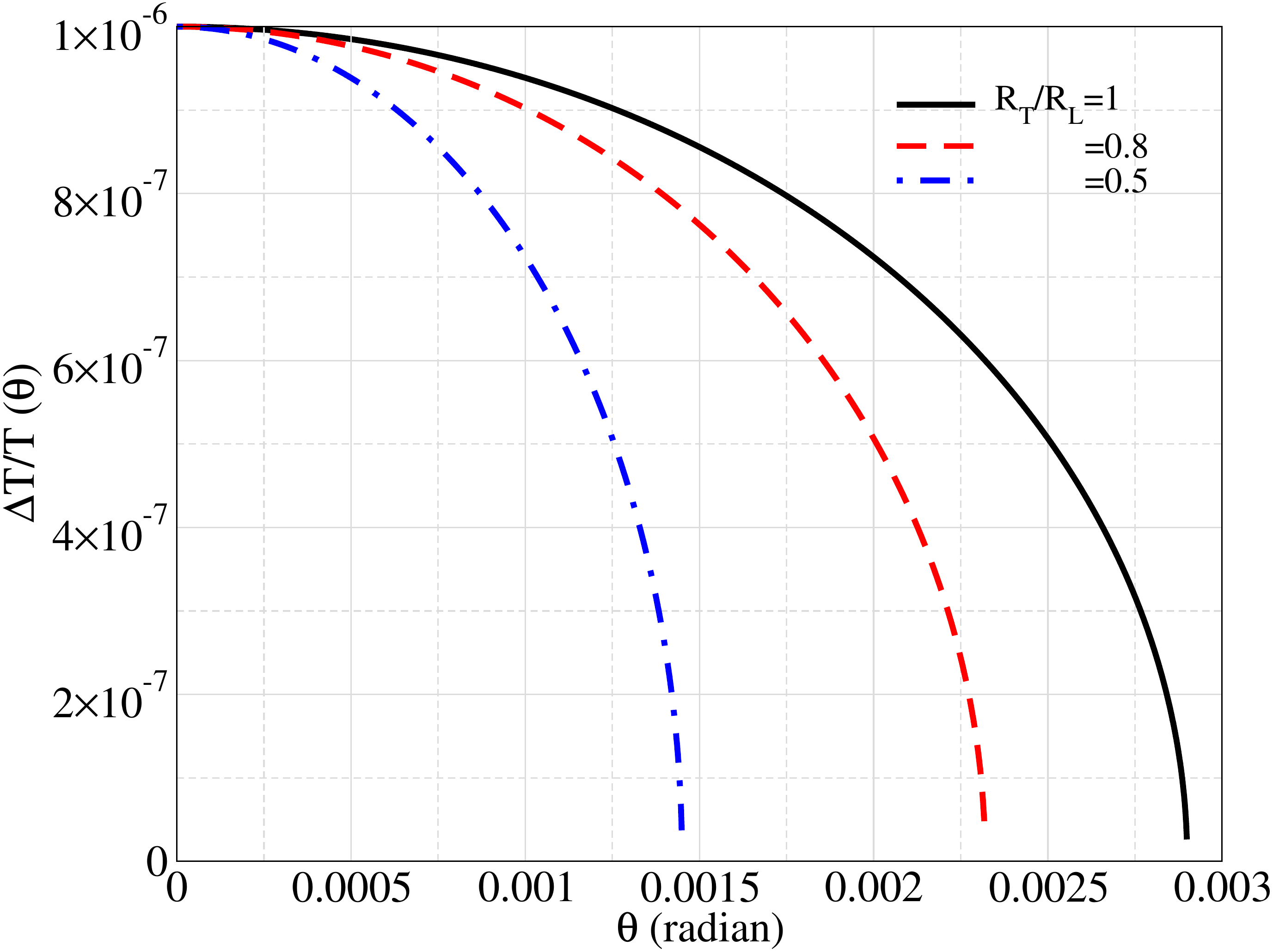}
\caption{kSZ temperature distortion for apparent non-spherical bubbles with different ratios for the transverse ($R_T$) to longitudinal ($R_L$) size of the bubble. The actual spherical bubble is located at $z=8$ with $\dot{N}=10^{58}$s$^{-1}, f_H=1$ and $t_Q\sim 10^7$ yr. We have used a higher $\dot{N}$ to get $\frac{\Delta T}{T}=10^{-6}$ which has a higher chance of detection (see text for details). 
}
\label{fig:temp_profile1}
\end{figure}

\begin{figure}
\centering 
\includegraphics[width=\columnwidth]{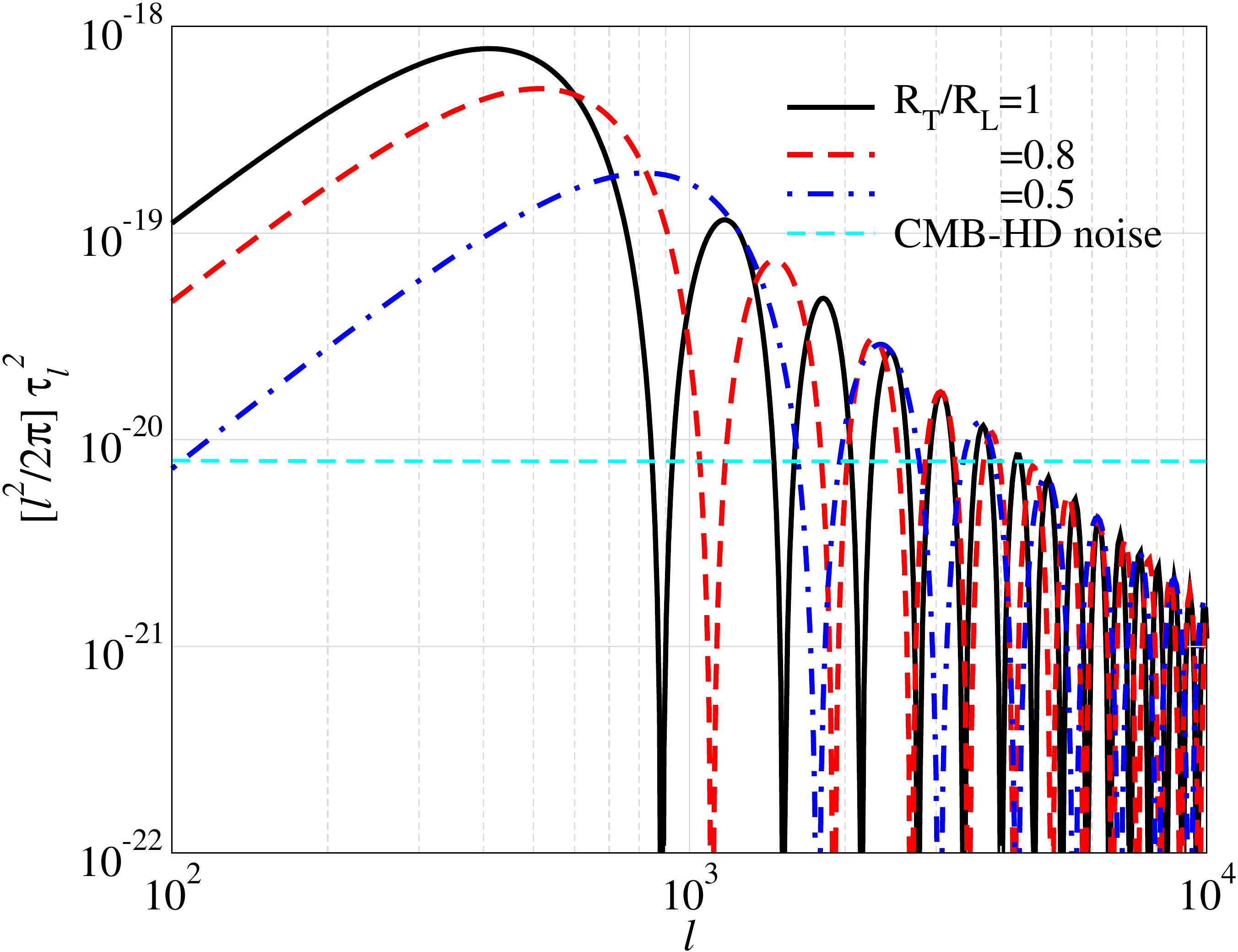}
\caption{Fourier transform of CMB anisotropy profiles of quasar bubbles for parameters as in Fig. \ref{fig:temp_profile1}
} 
\label{fig:temp_profile_fourier1}
\end{figure}

Up until now, our calculations assumed the observed bubbles to be spherical. However, the bubbles are expected to appear anisotropic to the observer since 
their measured sizes along the line of sight and transverse to it will have different observed values due to relativistic time delay \citep{WL2004,SIAA2006,Y2005,MBDC2011,MBC2012}. The size of the bubble in the transverse direction would appear to be smaller than the longitudinal direction due to a light travel time delay of $R/c$ (where $R$ is the size of bubble) compared to the age of the quasar $t_Q$. Thus, 
 the apparent transverse size $R_{\rm T}$ of the bubble is given by the size of bubble at time $t\,=\,t_Q-\frac{R}{c}$, whereas the longitudinal size $R_{\rm L}$ is given for $t\,=\,t_Q$.

\cite{Y2005} and \cite{MBC2012} have calculated the apparent distorted shape of a bubble due to relativistic expansion of the ionization front. The relationship between quasar age ($t_Q$) and its apparent age depends upon the angle of viewing and is given by,
\begin{equation}
t=t_Q-\frac{R(t)}{c}(1-cos(\phi)).
\end{equation}
For longitudinal direction, $\phi=0$ giving $t=t_Q$, while for transverse direction $\phi=\pi/2$ giving $t=t_Q\,-\,\frac{R(t)}{c}$. We can see from Fig. \ref{fig:distortion_shape} that at $t_Q=10^7$ yr (with $R\sim 2.4$ Mpc), the apparent age in transverse direction is $\sim 3\times 10^6$ yr. This causes the apparent transverse size of the bubble to be smaller than the longitudinal size by a factor of $\sim 3^{1/3}$. At $t\sim 10^8$ yr, the transverse size catches up with the longitudinal size as the bubble growth slows down or stops. 

In Fig. \ref{fig:temp_profile1}, we plot the temperature shift due to kSZ for a distorted bubble keeping the luminosity of quasar fixed for different $R_{\rm T}/R_{\rm L}$. This keeps $\left(\frac{\Delta T}{T}\right)_{\rm kSZ}$ through the center of the bubble fixed and changes the relative temperature profile of the bubble. It is clear that a greater shape-anisotropy  results is faster decline of the kSZ distortion profile. The resultant fourier imprint of the bubble for the different $R_{\rm T}/R_{\rm L}$ is shown in Fig. \ref{fig:temp_profile_fourier1}. Clearly, a highly distorted bubble shifts the oscillating features to higher multipoles which in turn gives a higher chance of detection since CMB fluctuations die out at high multipoles.

\begin{figure}
\centering 
\includegraphics[width=\columnwidth]{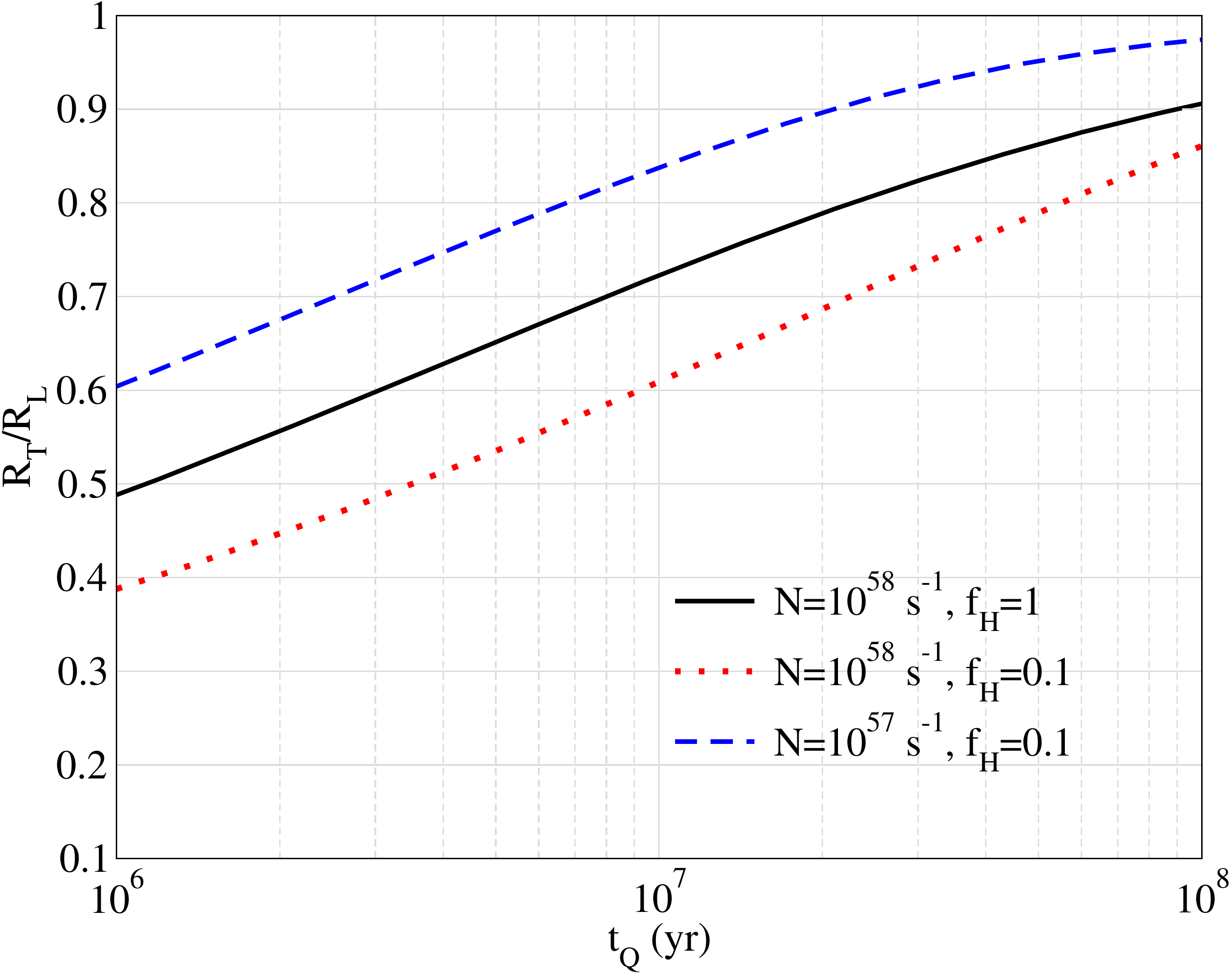}
\caption{Ratio of transverse to longitudinal size of a bubble as a function of $t_Q$ for few different $\dot{N}$. We consider a bubble in a neutral surrounding at $z=8$.
 }
\label{fig:RT_vs_RL}
\end{figure}

From a kSZ detection of a quasar bubble, one obtains the $y$-parameter ($\frac{\Delta T}{T}$) of the ionized bubble along with the ratio of transverse to longitudinal size of the bubble. 
The amplitude of the matched filter depends upon the longitudinal size ($R_L$) of the bubble as $y=n_e\sigma_T\frac{v_{rms}}{\sqrt{3}c}(2R_L)$. The inference of $R_L$, in turn, depends upon our knowledge of $n_e$ (or $f_H$).  The bulk linear velocity $v_{rms}$ can be estimated from a combination of CMB maps and galaxy surveys as was shown in \cite{ACT2016ksz,ACT2021ksz}. For a given $R_L$,  varying $R_T$ changes the Fourier profile of bubble (Fig. \ref{fig:temp_profile_fourier1}). Therefore, from the kSZ measurements, one obtains the knowledge of size ratio which is a function of $t_Q, f_H$ and $\dot{N}$ (Fig. \ref{fig:RT_vs_RL}). The detection of one bubble will not be enough to break this degeneracy and one would need multiple detections at a given redshift with quasars at different evolution stage. One can use 21 cm signal from the neutral IGM surrounding the bubble to break this degeneracy since ${\rm d}T_b\propto f_H$ \citep{WL2004}. Then, with the measurement of size ratio and $R_L$ from $y$, we can determine $t_Q$ and $\dot{N}$. 
One can also use a measurement of $f_H$ at a particular redshift from 21cm or Lyman-alpha emission observable to constrain these parameters. A detailed investigation of using complementary surveys in kSZ, Ly$_\alpha$ and 21cm to probe the bubble size, quasar physics and  bubble environment will presented in a future work.

\begin{figure}
\centering 
\includegraphics[width=\columnwidth]{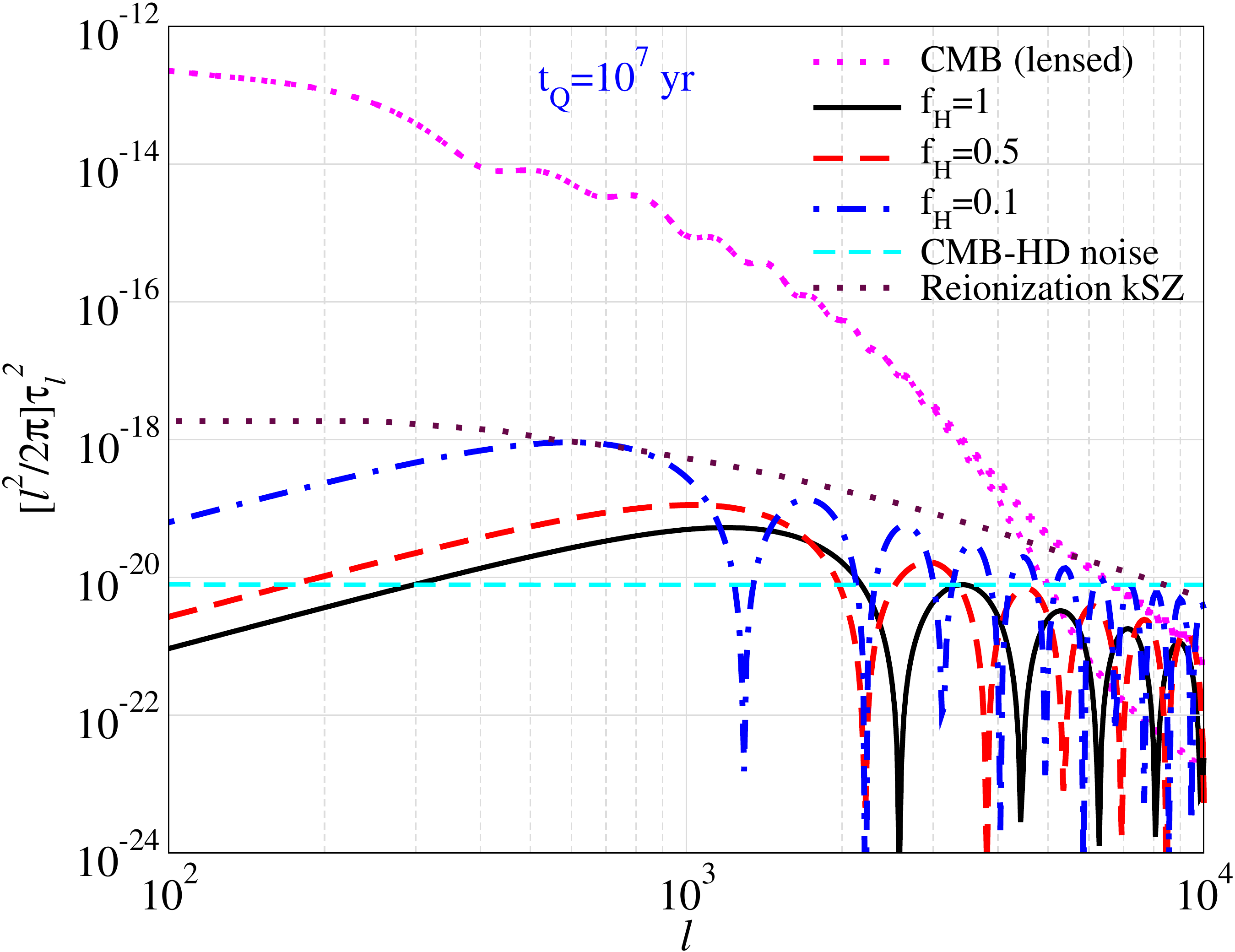}
\caption{Comparison of power spectrum of CMB and from quasar bubble for $t_Q=10^7$ yrs and at three different mean ionization fraction of hydrogen. We also show the noise from CMB-HD-like \citep{CMB_HD} experiment. Luminosity of quasar is $10^{58}$ s$^{-1}$ at $z=8$. For these choice of parameters, $R_T/R_L\sim 1/3$. 
}
\label{fig:matched_filter_t_1e7yr}
\end{figure}

\begin{figure}
\centering 
\includegraphics[width=\columnwidth]{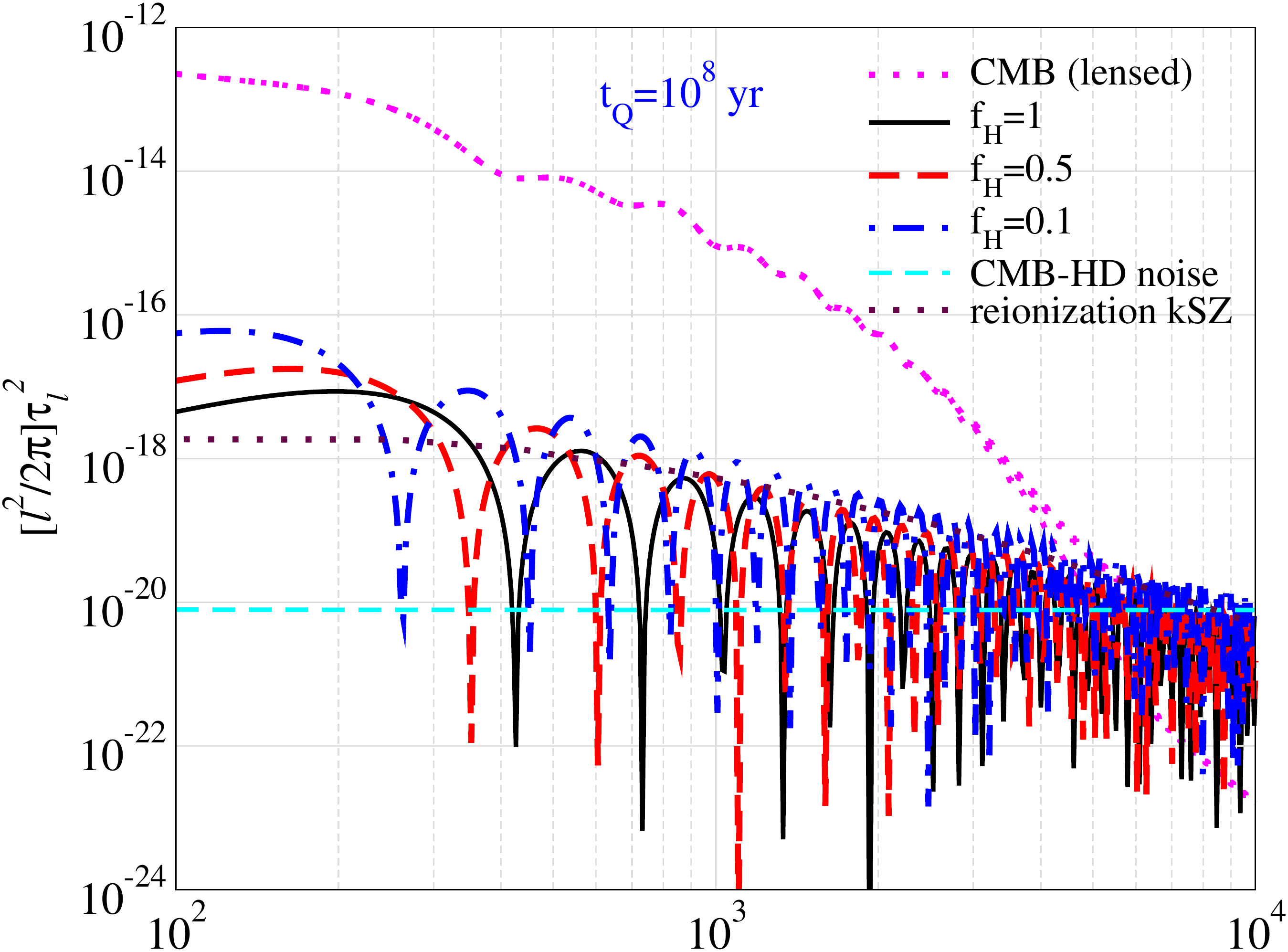}
\caption{Same as in Figure \ref{fig:matched_filter_t_1e7yr}, but $t_Q=10^8$ yrs. The corresponding $R_T/R_L\sim 1$. 
}
\label{fig:matched_filter_t_1e8yr}
\end{figure}

Since the signal to noise ratio depends upon the amplitude of the kSZ signal, bigger bubbles are likely to be detected first. Therefore, we consider a quasar with even a bigger $\dot{N}=10^{58}$s$^{-1}$. In Fig. \ref{fig:matched_filter_t_1e7yr}, we compare the power spectrum of this quasar bubble with CMB and the noise expected in the CMB-HD experiment. For $t_Q=10^7$ yr, the asphericity given by $R_L/R_T$ turns out to be $\sim 3$. The resulting $y$ for $f_H=1,0.5,0.1$ is $7.6\times 10^{-7}$, $9.5\times 10^{-7}$ and $1.6\times 10^{-6}$ respectively. This can lead to a detection of the bubble, with S/N at 0.17$\sigma$, 0.21$\sigma$ and 0.35 $\sigma$ respectively. We show the same for a larger $t_Q=10^8$ yr in Fig. \ref{fig:matched_filter_t_1e8yr}. In this case, the bubble size is even larger with $R_L/R_T\approx 1$ roughly equal to 1. This results in rapid oscillations of quasar bubble power spectrum (at high $\ell$) which makes it difficult to distinguish different scenarios. However, the bubbles are mostly spherical anyways and  the amplitude $y$ carries the relevant information. The value of $y$ for $f_H=0.1$ and 0.5 turns out to be $2.6\times 10^{-6}$ and $1.9\times 10^{-6}$ respectively. The resulting high S/N can lead to the bubble detection at 0.55$\sigma$ and 0.43$\sigma$, respectively. The same for $\dot{N}=10^{57}$s$^{-1}$ is equal to $1.2\times 10^{-6}$ and $8.9\times 10^{-7}$ which can be detected at 0.27$\sigma$ and 0.2$\sigma$ respectively.

\section{Conclusions}
In this work, we look at the secondary CMB anisotropies caused by the kinetic SZ effect  by ionized bubbles around quasars prior to reionization. In this regard, we model the kSZ distortion from a single quasar bubble and explore the possibility for its detection in future CMB experiments. We also calculate the resultant mean and rms temperature fluctuations of the CMB from a cosmological distribution of quasar bubbles.

The information gained from observing the kSZ imprint of a single quasar bubble depends ultimately on the search, and often serendipitous detection, of the quasar bubble either in kSZ or HI. It is easier, in principle, to search for the rms fluctuations from the quasar bubbles from a population of quasars at high redshifts.  This signal has been modeled in the past by \cite{ADPG1996}, who worked under the assumption of quasars being the prime source for the reionization of iur Universe.  
 We improve upon their work using a better prescription of the cosmological abundance of quasars in terms of the `halo-model'. We show that the kSZ power spectrum from quasars during reionization is negligible compared to the kSZ distortion from early galaxies found in simulations of reionization. As a direct consequence, we show that in our model, the ionized quasar bubbles cannot be responsible for the reionization of the Universe. However, we show that the amplitude and peak position of the quasar bubble temperature power spectrum depends on the reionization history, with a late reionization (say, $z_{re}\sim 6$)giving a higher signal by orders of magnitude compared to early reionization (say,  
  $z_{re}\sim 8$). Although, subdominant to the kSZ signal from late reionization due to galaxies, it might still be possible to extract the signal from the quasar bubbles if one can measure $C_{\ell}$ over multiple $\ell$-bins,  using the different $C_{\ell}$ templates to distinguish one source from other.

 We feel that a more promising approach would be to aim for individual quasar bubbles. However,  since the kSZ has a blackbody spectrum, and hence identical to the spectrum of primary CMB fluctuations,  it is not possible to distinguish kSZ from primary CMB  using spectral information alone. Notwithstanding, we can still selectively pickup the imprint of a quasar bubble on the CMB from our understanding of the very different spatial fluctuation of an ionized bubble w.rt the primary CMB. Especially, CMB fluctuations die away at smaller angular scales or higher $\ell$s due to diffusion damping while there is no strong dampening for quasar bubbles. Using 1-D radiative transfer calculations, we show that the quasar bubble has a hard boundary, i.e. box-like shape, which results in oscillating features in multipole space. The amplitude of the bubble signal and $\ell$-space ringing interval can be used to constrain a combination of quasar physics, the properties of the ambient medium and the bubble size and its velocity. Moreover, the 
 amplitude and shape of the kSZ signal depend upon the transverse versus the longitudinal size of bubble which may not be equal as the ionization front expands relativistically and we only see a snapshot of the quasar bubble on CMB. This gives an additional handle on physics behind the kSZ distortion template from a quasar bubble. A promising way to break the degeneracy between the intrinsic quasar parameters such as photon emission rate ($\dot{N}$) and lifetime ($t_Q$) and cosmological parameters like the large scale cosmological velocity and the neutral hydrogen fraction surrounding the quasar would be to 
 combine kSZ measurements, 21 cm and Lyman-alpha measurements from complementary surveys. The present work forms the stepping stone to a detailed investigation of the plausibility and importance of detecting individual quasar bubbles in kSZ will be presented in a future work.

\section*{Acknowledgements}
We would like to thank the referee for many important points which has helped us in raising the bar for this paper. We would also like to thank Geet Mankar in helping with making Fig. \ref{fig:fg_removal}.
 SKA was supported by the ERC Consolidator Grant {\it CMBSPEC} (No.~725456). SM acknowledges support of the Department of Atomic Energy, Government of India, under project no. 12-R\&D-TFR-5.02-0200. The authors acknowledge conversations with Sindhu Sri Sravya and Girish Kulkarni regarding reionization and bubble ionization profiles, and also for sharing their bubble profile results from simulations to which our analytical models were matched.

\section{Data availability}
The data are available upon reasonable request.

{\small
\vspace{-3mm}
\bibliographystyle{mn2e}
\bibliography{quasar}
}

\appendix
\section{Heating and cooling rates}
\label{app:rates}

In Fig. \ref{fig:cooling_rates}, we plot the cooling rates of various collision processes and photo-heating rate due to radiation from quasar. The cooling rates are function of temperature and number density of gas while photo-heating rate is proportional to $1/R^2$ when $R$ is the distance from quasar. In reality, the photo-heating rate is proportional to flux of photons which is $\propto \frac{e^{-\tau}}{R^2}$, where $\tau$ is the optical depth. If the photons are passing through ionized gas, there are no neutral hydrogen for photons to ionize and photon can free-stream to large distance. In that case, $\tau<<1$ and $e^{-\tau}\sim 1$. If there are neutral hydrogen, $\tau$ can be much greater than 1. Therefore, the photo-heating rate can be much smaller depending upon the number density of neutral hydrogen. In Fig. \ref{fig:cooling_rates}, we have assumed hydrogen to be ionized. The photo-ionization cross-section is given by \citep{ZS1989},
\begin{equation}
\sigma(E)=\frac{64\pi{\rm \sigma}_T}{\alpha^3}\left(\frac{I}{E}\right)^4\frac{{\rm exp}(-4\eta{\rm cot^{-1}}\eta)}{1-{\rm exp}(-2\pi\eta)},
\label{eq:photoionization}
\end{equation} 
where $E$ is the energy of photon, $I$ is the ionization energy of neutral hydrogen, $\alpha$=1/137, $\sigma_T$ is Thomson cross-section and $\eta=\frac{1}{\left((E/I)-1\right)^{1/2}}$.

{
\begin{figure}
\centering 
\includegraphics[width=\columnwidth]{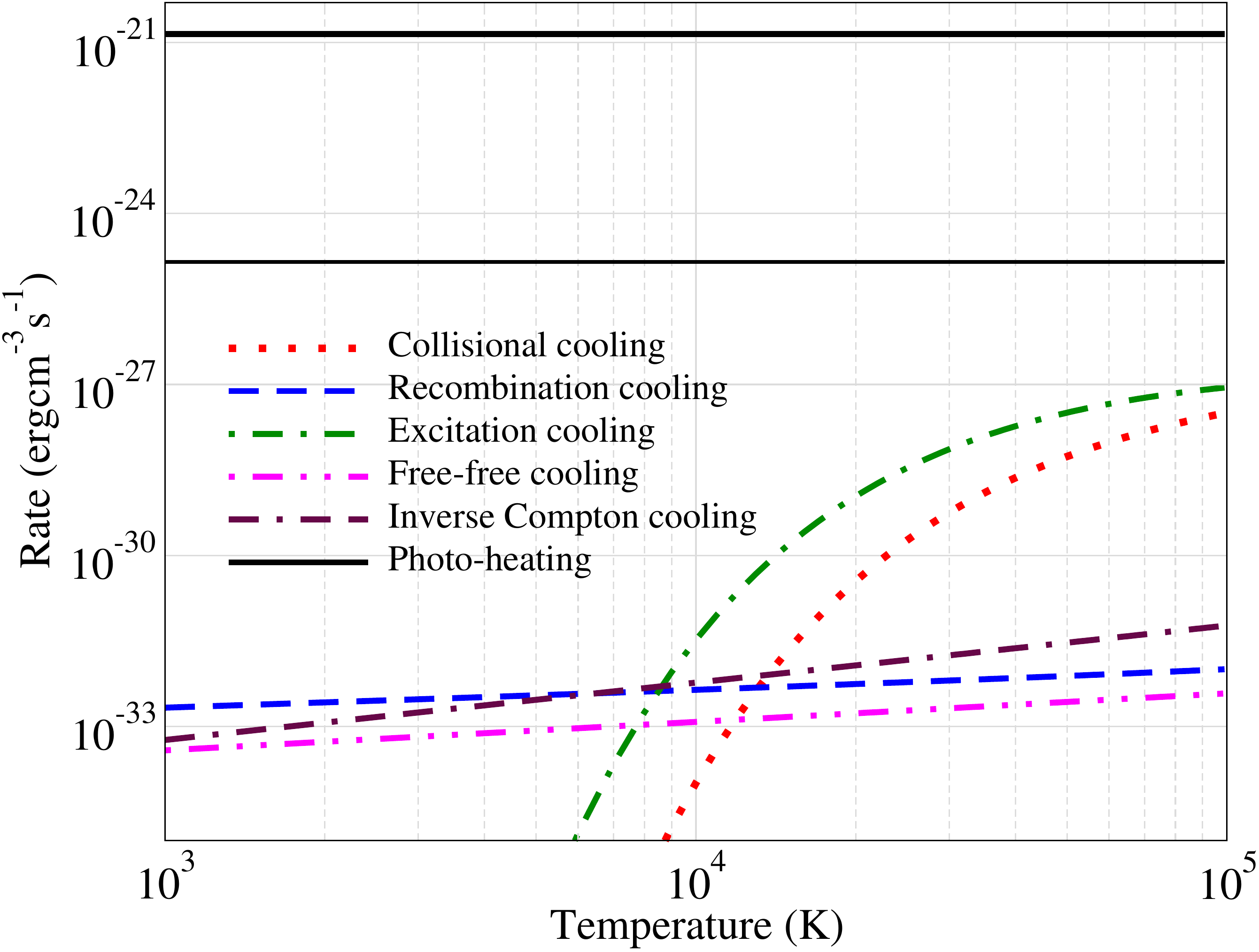}
\caption{Cooling rates as a function of temperature. As a comparison, we plot photo-heating rate due to the radiation at a distance 10 kpc (black thick solid) and 1000 kpc (black thin solid)  from the quasar. Ionization and excitation cooling decay exponentially below the corresponding energy threshold. We choose $\dot{N}=10^{57}$s$^{-1}$. The quasar is located at $z=6$. 
}  
\label{fig:cooling_rates}
\end{figure}
}

\end{document}